\documentclass[aps,prl,preprint,groupedaddress,longbibliography]{revtex4-1}
\usepackage{graphicx}
\usepackage{natbib}
\usepackage{amsmath,amssymb}
\usepackage{subfig}
\usepackage{setspace}
\usepackage{float}
\usepackage[normalem]{ulem}

\usepackage[usenames, dvipsnames]{color}
\usepackage{hyperref}
\hypersetup{
	hyperindex,
	breaklinks,
	colorlinks=true,
	linkcolor=blue,
	citecolor=magenta,
	bookmarks=true,
	bookmarksopen=true,
	bookmarksopenlevel=2,
	pdfstartpage={1},
	pdfstartview={FitH},
	pdfview={FitH 0},
	pdfauthor={B. F. Farrell and P. J. Ioannou},
	pdftitle={}}

\usepackage{ifthen}

\def\U{\bm{\mathsf{U}}}
\def\Uv{\mathbf{U}}
\def\uv{\mathbf{u}}

\def\U{\bm{\mathsf{U}}}

\def\Uv{\boldsymbol{U}}

\newcommand{\be}{\begin{equation}}
\newcommand{\ee}{\end{equation}}
\newcommand{\bdm}{\begin{equation*}}
\newcommand{\edm}{\end{equation*}}
\newcommand{\bea}{\begin{eqnarray}}
\newcommand{\eea}{\end{eqnarray}}

\newcommand{\partialf}[2]
{
 \ifthenelse{\equal{#1}{}}{\frac{\partial}{\partial #2}}{\frac{\partial #1}{\partial #2}}
}

\newcommand{\real}{\mathop{\mathrm{Re}}}

\renewcommand{\(}{\left(}
\renewcommand{\)}{\right)}

\newcommand{\upi}{\pi}

\newcommand{\df}{\textrm{d}}

\renewcommand{\i}{i}

\newcounter{saveeqn}%

\def\xv{\mathbf{x}}

\def\Uv{\mathbf{U}}
\def\uv{\mathbf{u}}

\newcommand{\defn}{\ensuremath{\stackrel{\mathrm{def}}{=}}}
\renewcommand{\equiv}{\defn}

\renewcommand{\U}{\mathbf{U}}
\renewcommand{\u}{\mathbf{u}}

\definecolor{fgreen}{rgb}{0.0, 0.5, 0.0}
\definecolor{dblue}{rgb}{0.2, 0.2, 0.6}
\definecolor{springgreen}{rgb}{0.09, 0.45, 0.27}
\definecolor{dartmouthgreen}{rgb}{0.05, 0.5, 0.06}
\definecolor{egyptianblue}{rgb}{0.06, 0.2, 0.65}
\definecolor{fireenginered}{rgb}{0.81, 0.09, 0.13}
\definecolor{forestgreen}{rgb}{0.0, 0.27, 0.13}
\definecolor{harvardcrimson}{rgb}{0.79, 0.0, 0.09}
\definecolor{amaranth}{rgb}{0.9, 0.17, 0.31}
\definecolor{mygreen}{rgb}{0,0.6,0}
\definecolor{mygray}{rgb}{0.5,0.5,0.5}
\definecolor{mymauve}{rgb}{0.58,0,0.82}

\begin{document}

\title{Statistical state dynamics-based study of the stability of the mean statistical state of wall-bounded turbulence}

\author{Brian F. Farrell}
\author{Petros J. Ioannou}
\affiliation{Department of Earth and Planetary Sciences, Harvard University, Cambridge MA 02138, USA}
\email{pjioannou@phys.uoa.gr}
%

\date{\today}

\begin{abstract}

Turbulence in wall-bounded flows is characterized by stable statistics for the mean flow 
and the fluctuations both for the case of the ensemble and the time mean. Although, in a substantial  set of turbulent systems, this stable statistical 
state corresponds to a stable fixed point  of an associated statistical state dynamics (SSD) 
closed at second order, 
referred to as S3T,
this is not the case for wall-turbulence.   In wall-turbulence 
the trajectory of the statistical state evolves on a transient chaotic  attractor in the S3T statistical state phase space  and
the time-mean statistical state is neither a stable fixed point  of this SSD nor, if 
the time-mean statistical state is maintained as an equilibrium state, is it  stable.
Nevertheless, sufficiently small perturbations from the ensemble/time-mean state of wall-turbulence
are expected  to relax back  to the mean statistical state  following an effective linear dynamics.  
 In this work the dynamics
of spanwise uniform perturbations to the time-mean flow  are studied using  a linear inverse model (LIM) to identify 
the linear  operator governing the ensemble stability  of the ensemble/time-mean state by obtaining the time mean stability properties over the transient 
attractor of the turbulence identified by the S3T SSD.
The ensemble/time-mean stability of an  unstable equilibrium
can be understood by 
noting that even when every member of an ensemble is unstable the ensemble mean may be  stable with 
perturbations following an identifiable stable dynamics.  While simplifying insight into turbulent flows has commonly 
been obtained by identifying and studying  ensemble mean  statistical states, 
less attention has been accorded to identifying and studying the ensemble mean dynamics.
We show that in the case of
 wall turbulence, even though stable 
fixed point SSD equilibria  are not available to allow application of traditional perturbation analysis
methods to identify the perturbation stability of the mean state,
an effective linear stability  analysis  can be obtained 
to identify the perturbation dynamics of the ensemble/time-mean statistical state.
 \end{abstract}

\maketitle

\section{Introduction}



{\color{black} Turbulence in parallel channel Couette flow (pCf) and pipe Poiseuille flow
(pPf) at low Reynolds numbers lies  on a transient chaotic attractor  in state space \citep{Brosa-89,Eckhardt-etal-2007}.
However, even in the low  Reynolds number simulations  used  in this work, turbulence persists  long enough  so that the  first two moments of the  statistics 
required by our analysis are converged.}
The time-mean statistics of the  turbulent state in  plane Couette flow (pCf) and plane Poiseuille flow (pPf) 
comprises the time-mean flow $\langle \U \rangle= U(y) \hat{\xv}$, which is confined to  the streamwise direction, $x$, and 
 depends
only on the cross-stream direction, $y$, together with the cumulants of fluctuations from this statistical state, 
which are homogeneous functions of the streamwise and spanwise, $z$, direction.
This statistical mean turbulent state  depends only on the Reynolds number   and is
stable in the sense that almost any  perturbation
to the flow will  result in the flow relaxing back to its original statistics.
If perturbations to the statistical mean state are sufficiently small,  the dynamics of these perturbations is expected to be 
linear and therefore controlled by a linear operator characterized by its eigenmodes and eigenvalues.
Recently, the dominant eigenmodes and eigenvalues of the linear operator underlying the dynamics of perturbations to 
the time-mean velocity profile in turbulent pPf were estimated empirically using a DNS ensemble by  
\citet{Iyer-etal-2019} (hereafter referred to as IWCV).
IWCV succeeded in estimating the dominant eigenvalues, which characterize the dynamics  of relaxation to the time-mean statistical state of all the cumulants of the statistical state,
and also the dominant mean-flow eigenmodes, which characterize the least-damped  perturbation structures producing relaxation to the time-mean flow.

%

While statistical stability in the sense  of the existence of a stable linear dynamics underlying the return of a perturbed statistical state
back to its  stationary statistics is expected, and the streamwise and spanwise mean flow component of the first two least stable eigenmodes 
for the time-mean velocity profile have been estimated,
identifying the  dynamics of the statistical state
stability and its physical mechanism would seem to require  solving the  statistical state dynamics (SSD) for the full statistical state comprising both
the mean state and  the higher order cumulants of
the fluctuations, which in the standard ensemble formulation of  SSD 
would  require
obtaining the dynamics of all the cumulants. 
Remarkably,  this program  can be accomplished, to the degree that  Gaussian statistics govern the essential dynamics
of the SSD,  by solving  only for the first two cumulants in the appropriate SSD. Using this SSD, referred to as S3T,
the statistical stability of the ensemble/time-mean state has been analytically determined in
a  diverse class of turbulent flows 
which share  the property that the statistical  state is attracted to a stable fixed point equilibrium.
The success of this program is predicated on the choice of the averaging operator  underlying the S3T SSD.
It is important that this averaging operator be the streamwise average,
rather than the ensemble average in order that the minimal nontrivial  SSD, in which the expansion in cumulants is closed
at second order, corresponds to the mechanism of the turbulence.
 An important attribute of the S3T minimal mechanistically complete SSD is that it provides a Gaussian approximation 
 for all the fluctuation statistics so that, to 
 the extent that Gaussian statistics underlie the fundamental dynamics 
 maintaining wall-turbulence, which has been verified \cite{Bretheim-etal-2015,Farrell-etal-2016-VLSM},  understanding mechanism in the transparently 
 simple S3T SSD is tantamount to understanding wall-turbulence.
%

While  turbulent systems with stationary statistics are characterized by a statistical equilibrium state 
to which ensemble SSD  trajectories converge,
the  corresponding  S3T SSD  trajectory may identify an underlying transient chaotic attractor, with the 
dynamics of the approach to the statistical mean state resulting from averaging over this fundamental attractor 
of the turbulence.  
 Among the turbulent systems for which  averaging over the S3T attractor is not required,
 because the same fixed point equilibrium is obtained in both  the ensemble and  the S3T  SSD, 
 are 2D 
$\beta$-plane turbulence \citep{Constantinou-etal-2016}, 3D baroclinic turbulence \citep{Farrell-Ioannou-2008-baroclinic,Bakas-Ioannou-2019-book},  drift-wave turbulence in plasmas \citep{Farrell-Ioannou-2009-plasmas,Constantinou-Parker-2018}, and  {\color{black} wall-bounded shear flow 
 in the presence of free-stream fluctuations but before transition to self-sustaining turbulence 
 (referred to as pre-transitional flow)}\citep{Farrell-Ioannou-2012,Farrell-Ioannou-2017-bifur}.
In these cases  the   stability  of the S3T time-mean state, which coincides 
with the ensemble mean state, can be determined by eigenanalysis of  the S3T operator 
perturbed about the time-mean state. 
{\color{black} In wall-bounded flows 
after transition to turbulence (referred to as post-transitional flow)  no stable fixed points 
of the S3T SSD exist  and the S3T state trajectory lies on the fundamental  transient chaotic 
attractor of wall-turbulence that it identifies \citep{Farrell-Ioannou-2012,Farrell-Ioannou-2017-bifur}.} Therefore, the program 
of analytically determining the stability of the statistical state of a post-transitional  turbulent shear flow 
by finding and perturbing the  fixed point of its  S3T SSD 
equations fails. It is useful to comment on  some implications of this result:

\begin{enumerate}
\item     The instability of the ensemble/time mean state 
identified in the S3T SSD is a physical instability  that can be observed in N-S turbulence.  
It is a member of  a set of nonlinear instabilities resulting from interaction between the mean shear flow and fluctuations to the 
mean shear flow. S3T   allows analytic expression to be obtained 
for this novel class of instabilities in turbulent shear flow \cite{Farrell-Ioannou-2003-structural,Farrell-Ioannou-2012,Srinivasan-Young-2012,Parker-Krommes-2013,Bakas-Ioannou-2013-prl,Bakas-Ioannou-2013-jas,Srinivasan-Young-2014,Parker-Krommes-2014-generation,Constantinou-etal-2014,Bakas-etal-2015,Constantinou-etal-2016,Farrell-Ioannou-2017-bifur,Farrell-Ioannou-2017-Saturn}.

\item There is no contradiction arising from the required stability of the
ensemble/time mean state  in ensemble dynamics and its  instability when considered in S3T dynamics. 
Stability of the ensemble mean state is  implied by   stationarity of the statistics of the turbulent flow
and can occur even when each member of the ensemble  is unstable    as consideration of the stochastically modulated Mathieu 
equation governing the ensemble evolution of the parametric mass-spring example shows.
In that example every 
ensemble member is unstable while the ensemble mean has stable limit cycle dynamics \cite{Farrell-Ioannou-2002-perturbation-I}. 

\item  The lack of a stable fixed point in S3T dynamics corresponding to the ensemble/time-mean flow implies the non-existence  of a point attractor.
 In these cases the pertinent  dynamics  needs to be obtained by  ensembling over the transient chaotic attractor.
 The emergent  dynamics obtained by averaging  over this attractor can be an object of study analogous to studying the ensemble mean statistics
 of the turbulent state. 
For example, consider that a covariance matrix and a lag covariance matrix 
 have been obtained from observations of a turbulent flow.  Each of these is a statistical quantity and the object of traditional study
 by ensemble methods.   However, the mapping between them is a linear operator related directly to 
 the dynamics of the turbulence rather than to its statistical state.  
 It is this emergent ensemble dynamical object that provides new insight into turbulence.

\item The dynamics of the S3T SSD  is chosen explicitly to isolate the essential components of turbulence dynamics: the mean flow,  the second cumulant
and the interaction between them.  In contrast to the familiar 
concept of the turbulent state as a point moving along a chaotic trajectory lying on an attractor embedded in the state space of velocity  \cite{Keefe-etal-1992}, 
the SSD viewpoint, which makes possible fundamental dynamical insight,
is of turbulence as a point  moving along a chaotic trajectory in the SSD state space which consists of the mean flow and the 
higher  velocity cumulants. Remarkably, the essential dynamics of wall-turbulence is obtained using the simplest 
nontrivial closure of the SSD which is to retain only the first and second cumulants. It is useful for visualization purposes to translate 
the state space of this second order closure of the SSD
to velocity variables  in which  the mean flow point  (first cumulant) and its surrounding probability distribution  of fluctuations (second cumulant)
follows a  
chaotic trajectory. 
The  probability distribution  obtained 
from the covariance matrix of the second cumulant in  S3T SSD 
is multivariate Gaussian  while the exact distribution, which can be obtained 
by closing with the  third cumulant  from a DNS, is slightly non-Gaussian. The iso-density locus of the multivariate Gaussian 
probability density function (PDF) forms an elliptical distribution. 
The axes and orientation of this elliptical distribution determine the Reynolds stresses that, together with the nonlinear mean flow dynamics, determine the 
chaotic trajectory  in the state space of the SSD.  
Adopting this SSD state space chaotic attractor viewpoint of turbulence 
dynamics is motivated further by the observation that wall-turbulence at high Reynolds numbers can be regarded as a covering (tiling)
of the turbulent channel with minimal channel units \cite{Jimenez-Moin-1991,Flores-Jimenez-2010,Jimenez-Kawahara-2013},  the dynamics of each of which is closely approximated by a minimal 
channel S3T SSD, to form an ensemble covering of the attractor that would have ensemble mean dynamics equivalent to the time 
mean dynamics obtained by time integration over one of these units.  A related study recently showed that displacing the 
observed minimal channel tiles in a DNS of wall-turbulence, so that the roll-streak structure in the tiles is aligned in the streamwise 
direction,  recovered the S3T dynamics. Given that the roll-streak structures so aligned have wavenumber zero in the streamwise direction, as required in the S3T formalism,
verifies that the S3T SSD dynamics, which is associated with an analytically characterized transient chaotic attractor in SSD state space,
 is also relevant to understanding wall-turbulence at high Reynolds number \cite{Nikolaidis-POD-2023,Nikolaidis-RS-2023}.
 
\item While there are differences in the statistical distributions obtained between DNS 
and S3T \cite{Hwang-Eck-2020,Hernandez-Hwang-2020,Hernandez-Hwang-2022}, these differences arise in conjunction with the simplification of the dynamics of S3T SSD that allows 
detailed analysis  of the mechanism underlying the turbulence and the fact that these differences do not affect the fundamental dynamics (e.g. SSP cycle, wall stresses,
mean velocities \cite{Bretheim-etal-2015,Farrell-etal-2016-VLSM,Bretheim-etal-2018}) indicates that
 these differences are inessential.


\end{enumerate}

{\color{black} It should be additionally noted that  the attractor of the S3T SSD in 
cumulant variables differs from the attractor   of the corresponding turbulent state  represented in  its velocity variables.
For example, in beta-plane turbulence  the turbulent state  in velocity variables follows a chaotic trajectory, while
in the corresponding S3T SSD expressed in its statistical state cumulant variables the attractor   is  most often 
a  stable fixed point \cite{Farrell-Ioannou-2003-structural}.
Post-transitional wall-turbulence presents a case in which both the 
S3T SSD in cumulant variables  and the turbulent state in velocity variables lie on distinct
transient chaotic attractors. 
It follows that  the  ensemble stability properties of the  S3T SSD can be obtained by invoking ergodicity and averaging over the S3T SSD
 chaotic trajectory. An alternative is to exploit ergodicity by seeding an ensemble
of perturbations over the reflection of the S3T attractor in DNS in order to explore its ensemble  stability properties, as in  the analysis of  IWCV.
Agreement between these very different conceptual and computational approaches lends credence to the view 
of an emergent ensemble 
stability dynamics arising as an average dynamics, the average being taken over an attractor whether it be through ensemble or  time averaging.  
It also lends support to viewing turbulence as lying on an attractor in statistical state space distinct from the traditional attractor in velocity space \cite{Keefe-etal-1992}.

  }

In order to properly interpret   stability analysis applied to statistical states, it is important to distinguish  the instability
of an SSD state in cumulant variables  from the more familiar 
hydrodynamic instability  of a flow state in velocity variables.
In  linear hydrodynamic stability studies an unperturbed flow state   is maintained while   growth or decay of perturbations to this  flow is examined. 
For example,   in the absence of fluctuations,   laminar  pCf is an equilibrium state and the associated linear 
perturbation equation has eigenvalues with negative real part indicating the pCf is hydrodynamically 
stable to infinitesimal perturbations at all Reynolds numbers.
By contrast, 
SSD stability examines the stability of the  cumulants, which are the state variables of the SSD equations
(cf. \cite{Farrell-Ioannou-2003-structural,Farrell-Ioannou-2019-book,Markeviciute-Kerswell_2023}).
 Linear SSD stability analysis subsumes  linear hydrodynamic stability analysis: in the absence
of background fluctuations producing non-vanishing higher order cumulants,  the hydrodynamic stability of the associated laminar flow assures  also its  SSD stability.
However, the SSD equations may also  support additional instabilities when the flow state
contains a non-vanishing second order cumulant.
These instabilities arise from the interaction between  perturbations to the  mean flow and the second cumulant of the unperturbed SSD state, which is absent
in hydrodynamic stability analysis. 
This interaction is familiar as it underlies
 the self-sustaining process in wall-bounded flows \cite{Hamilton-etal-1995,Waleffe-1997}.
It is important to note that the S3T-SSD,   which incorporates quadratic variables,  allows us to determine using  linear eigenanalysis 
the existence of this set of nonlinear instabilities supported by
the Navier-Stokes equations in wall-bounded shear flow. 
%
%
An example of such a nonlinear instability arises in
stationary spanwise independent mean flow equilibria  maintained by 
a spanwise homogeneous field of  turbulent fluctuations in the S3T SSD of pCf.
These spanwise independent mean flow equilibria are SSD unstable for sufficiently high 
Reynolds number and in the presence of sufficient  stochastically maintained turbulence
both in the framework of the S3T SSD \cite{Farrell-Ioannou-2012}, and 
by DNS ensemble approximations to an SSD closed at infinite order \cite{Farrell-Ioannou-2017-bifur}.
The unstable modes that 
arise from SSD instability of these spanwise independent SSD mean flow equilibria
have the form of streamwise roll-streak structures
that break the spanwise homogeneity of the 
streamwise and spanwise independent SSD equilibrium state.
Importantly, over a range of Reynolds numbers and levels of stochastically maintained turbulence, these instabilities equilibrate 
to form stable finite amplitude fixed point states with roll-streak (R-S) structure.  
The stability of these R-S states was verified 
by study of the perturbation dynamics of these SSD equilibria \cite{Farrell-Ioannou-2017-bifur}.   
However, for high enough Reynolds numbers and levels of stochastically excited turbulence, there 
is no stable fixed point equilibrium statistical state and the statistical state of the turbulence lies on a transient chaotic attractor.  
Turbulence, once established on  this attractor of the statistical state, 
continues to be maintained when the stochastic excitation 
responsible for its inception is removed, indicating that the turbulence is 
self-maintained absent typically rare relaminarization events \citep{Farrell-Ioannou-2012,Farrell-etal-2016-VLSM}. 
In this work we show, within the framework of
the S3T SSD,  that
the time-mean flow that is  self-maintained in turbulent pCf is S3T-SSD unstable with  eigenmodes 
in the form of streamwise R-S structures together with  supporting perturbations in the form of the second order cumulant.
%

At this point in our study we will have  verified that the statistical mean state of turbulent pCf  is neither a state of marginal 
hydrodynamic stability nor is it a state of 
S3T SSD stability,  rather it is SSD unstable  to structures with R-S form, 
which leaves open the question of explaining and quantifying the observed statistical stability of the time-mean flow in pCf.
That a physical instability of the time-mean state is verified  in this paper to exist in the S3T SSD 
suffices to ensure that this mean flow is not a fixed point about which 
traditional time independent stability analysis to obtain 
eigenmodes and eigenvectors can be applied.  
Nevertheless, perturbation  stability of any SSD with stationary statistics is expected.
The first step in analysis of the origin and nature of the expected linear stability of a time-mean flow would be to obtain the eigenvalues and
eigenmodes  of the necessarily linear dynamics of streamwise and spanwise constant perturbations to this  time-mean flow.
In addressing this question in pPf turbulence, IWCV used an ensemble method  to obtain empirically the first 
two eigenvalues and eigenmodes of this  linear dynamics.
An open question is what this stable empirical linear dynamics  represents. 
 To address this question we have obtained an effective linear dynamics
governing perturbations to the time-mean statistical state   in a quasi-linear pCf, for which we have extensive analytic characterization,
 and in  a pPf DNS,  which extends the results of the analytically characterized SSD  and also makes connection with IWCV. The effective 
 linear dynamics was obtained
 using the linear inverse model (LIM) method,  which has been applied widely in geophysical fluid dynamics 
\cite{Penland-1989,Penland-Ghil-1993,Penland-1995a,DelSole-04}.  
An early application of LIM was to diagnose the mechanism and predict 
the evolution of the El-Nino Souther Oscillation in the Tropical Pacific \cite{Penland-Magorian-1993,Penland-1995a}.
 In addition, LIM was used to 
determine  the effective dynamics governing the climate statistics of an  
atmospheric  model \cite{DelSole-Hou-1999}  and the low-frequency variability of the midlatitude climate  \cite{Winkler-etal-2001}.
In another application, LIM analysis was used to show that
that the eddy stresses interacting with the mean flow in two-layer quasi-geostrophic 
turbulence can be diagnosed  to comprise the action of
upgradient momentun transport together with eddy-diffusion  and  stochastic excitation   \cite{DelSole-Farrell-1996}.  
In an early application to fluid dynamics LIM was used to show that the dynamics of a dilute gas in a Rayleigh-B\'enard configuration near criticality
reproduces the linearized N-S equations excited with  stochastic forcing with the covariance  predicted by the 
Landau-Lifschitz theory \cite{Penland-Garcia-1991}.

LIM analysis infers the dynamics underlying a temporal sequence of simulation data
from the covariance of the data  and the time advanced covariance exploiting
ergodicity to
interpret the linear dynamics obtained 
as the ensemble linear dynamics with the ensemble taken over the transient chaotic  attractor of the turbulent state. 
In both the LIM time averaged methods and the IWCV ensemble method,
 the 
 eigenvalues and eigenmodes of perturbations from the time-mean flow  correspond to 
the temporal evolution and structure of the least damped modes governing return to the
stable stationary fixed point  of the ensemble SSD.

{\color{black}
It is instructive to note that more generally LIM analysis  addresses the issue of providing
the optimal linear mean operator and excitation 
for  resolvent analysis. In resolvent analysis the operator of the linear dynamics is prescribed
to be the operator that governs the evolution of perturbations about the time-mean, possibly modified by eddy viscosity
\cite{Sharma-McKeon-2013,McKeon-2017,Hwang-Eck-2020}, and the accuracy of the predictions of resolvent
analysis is predicated on the  structure and spectrum of the
input excitations, which is a topic being actively researched (cf. \citet{Zare-etal-2017,Towne-Lozano-2020,Bae-etal-2021, Morra-Henningson-2021,Holford-etal-2023,Holford-Hwang-2023,Abootorabi-Zare-2023}). 
LIM obtains the Langevin form of the ensemble perturbation dynamics that resolvent analysis relies upon
including both 
the operator of the linear dynamics and  the covariance of the associated excitation. From this perspective LIM can be viewed as providing a method for constructing 
a linear  model for the dynamics of fluctuations in turbulence in which the ambiguity in the structure of the excitation has been resolved.}

We conclude that in  post-transitional wall-turbulence no  stable 
fixed point  exists that would correspond to 
the stable fixed points of the S3T SSD in the pre-transitional turbulent state, which 
allowed the modes to be identified directly by perturbing the SSD dynamics linearized about this stable stationary point.
However, nothing essential is lost, insofar as the dynamics of  perturbations to the time-mean flow is concerned,
as the LIM and ensemble methods both allow the effective linear dynamics 
of perturbations to the time-mean flow averaged over the fundamental structure of the transient chaotic attractor to be identified.

\section{Formulation of   the S3T SSD stability analysis for wall-turbulence}

Consider a pCf with streamwise direction, $x$, wall-normal direction, $y$, and spanwise direction, $z$.
The  lengths of the
channel in the streamwise, wall-normal and spanwise  direction are respectively $L_x$, $2h$ and $L_z$. 
The channel walls are at $y/h=-1$~and~$1$.  
Averages are denoted by  angle brackets with a subscript denoting the independent variable over which the average is taken, i.e.~streamwise averages by $\langle \,\boldsymbol{\cdot}\,\rangle_x=L_x^{-1} \int_0^{L_x} \boldsymbol{\cdot}\ \df x$,
time averages by $\langle \,\boldsymbol{\cdot}\,\rangle_t=T^{-1} \int_0^{T} \boldsymbol{\cdot}\ \df t$,  
and 
ensemble averages  over different realizations of the flow obtained from different initial conditions by  $\left < \cdot \right >_{E}$. 
In order to proceed with the formulation of the S3T SSD closed at second order we  choose as an averaging operator  the streamwise mean.
 This is a crucial choice, because it allows the formulation of a second order mean field theory that supports realistic turbulence, including in pCf 
and pPf \citep{Farrell-Ioannou-2012,Farrell-etal-2016-VLSM}.  In this closure the vector velocity, $\uv$, is  decomposed into its streamwise mean, denoted by $\Uv(y,z,t)$
and the deviation from this mean (the fluctuations) 
denoted $\u'(x,y,z,t)$ so that  $\uv = \Uv + \uv'$. The pressure gradient is similarly decomposed as $\nabla p= \nabla\( P(y,z,t)+ p'(x,y,z,t) \)$.  Velocity is  non-dimensionalized
by the velocity at the wall, $U_w$, at $y/h=1$, lengths by $h$, and time by $h/U_w$.   The non-dimensional NS equations decomposed into an equation for the  mean and an equation for the 
fluctuations are:
\begin{subequations}
\label{eq:NS}
\begin{gather}
\partial_t\mathbf{U}+ \mathbf{U} \cdot \nabla \mathbf{U}   + \nabla P -  \Delta \mathbf{U}/R = - \langle \mathbf{u}' \cdot \nabla \mathbf{u}'  \rangle_x\ ,
\label{eq:NSm}\\
 \partial_t\mathbf{u}'+   \mathbf{U} \cdot \nabla \mathbf{u}' +
\mathbf{u}' \cdot \nabla \mathbf{U}  + \nabla {p}' -  \Delta  \mathbf{u}'/R
= - \(  \mathbf{u}' \cdot \nabla \mathbf{u}' - \langle \mathbf{u}' \cdot \nabla \mathbf{u}'\rangle_x \,\) \ ,
 \label{eq:NSp}\\
 \nabla \cdot \mathbf{U} = 0\ ,\ \ \ \nabla \cdot \mathbf{u}' = 0\ . \label{eq:NSdiv0}
\end{gather}\label{eq:NSE0}\end{subequations}
where $R= U_w h/ \nu$ is the Reynolds number.
The velocities satisfy periodic boundary conditions  in the $z$ and $x$ directions
and no-slip boundary conditions in the cross-stream direction: $\mathbf{U}(x,\pm 1,z,t)= (\pm 1,0,0)$, $\mathbf{u}'(x,\pm1,z,t)=0$.

The S3T SSD, which is based on the crucial choice of a streamwise mean for the averaging operator, 
when closed at second order embodies the fundamental dynamics underlying wall-turbulence.  
It is the S3T SSD that allows analytic solutions to be found and makes direct connection to 
canonical cumulant expansion methods and insights.   A  highly
accurate second order cumulant approximation to the 
S3T SSD that provides motivation for deriving the S3T SSD as well as  a powerful computational tool
can be directly obtained from the quasilinear approximation of the NS equations \eqref{eq:NS}:  
\begin{subequations}
\label{eq:RNL}
\begin{gather}
\partial_t\mathbf{U}+ \mathbf{U} \cdot \nabla \mathbf{U}   + \nabla P -  \Delta \mathbf{U}/R = - \langle \mathbf{u}' \cdot \nabla \mathbf{u}'  \rangle_x\ ,
\label{eq:RNLm}\\
 \partial_t\mathbf{u}'+   \mathbf{U} \cdot \nabla \mathbf{u}' +
\mathbf{u}' \cdot \nabla \mathbf{U}  + \nabla {p}' -  \Delta  \mathbf{u}'/R
= 0\ ,
 \label{eq:RNLp}\\
 \nabla \cdot \mathbf{U} = 0\ ,\ \ \ \nabla \cdot \mathbf{u}' = 0\ . \label{eq:RNLdiv0}
\end{gather}\label{eq:RNL00}
\end{subequations}
which entails neglecting or parameterizing the fluctuation-fluctuation interactions in 
\eqref{eq:NSp} while retaining the fluctuation-fluctuation interactions in \eqref{eq:NSm} (cf. \cite{Farrell-Ioannou-2012,Marston-2012,Marston-2023,Markeviciute-Kerswell_2023}). 
Here we neglect altogether the fluctuation-fluctuation interactions in  \eqref{eq:NSp}. 
Neglecting altogether the fluctuation-fluctuation interactions in \eqref{eq:NSp} has no
fundamental effect on the 
turbulence in the sense that the turbulent state is supported with the mean and integral scales as well as the energy 
extracting scales of the fluctuations  being similar    to those of a DNS of pCf. This quasi-linear system, which approximates the S3T SSD, 
is referred to as the restricted
non-linear system (RNL)  (cf. \cite{Thomas-etal-2014,Thomas-etal-2015,Bretheim-etal-2015,Farrell-etal-2016-PTRSA}).
{\color{black} It is worth noting that 
the underlying dynamical structure of the quasi-linear equations has been verified
by \citet{Alizard-2017,Alizard-2019,Pausch-etal-2019} to support  edge states and to have the bifurcation behavior
of exact coherent structures of the Navier-Stokes equations.}

The  S3T SSD  describes the composite dynamics resulting from  the interaction of  an ensemble
of fluctuations, each of which evolves under the same streamwise-mean flow $\U$, with the streamwise-mean flow.  
This choice of adopting the streamwise average in the  S3T SSD formulation is physically motivated: 
this SSD captures the essential dynamics of the turbulence at second order, which are the self-sustaining process and its regulation. 
The variables   of the S3T SSD  are 
the first two cumulants consisting of the
streamwise mean flow,
$\U=(U,V,W)$  or $\U  \equiv (U_x,U_y,U_z)$,
and the second order cumulants that are the same time ensemble mean covariances of the Fourier components of the  velocity fluctuations, $\hat{u}_{\alpha, k_x}'$,
where the index $\alpha=x,y,z$ indicates the velocity  component in the Fourier expansion
of the  perturbation velocity  $\uv'$ :
\begin{equation}
\uv' (x,y,z,t)= \sum_{k_x>0} \real \left (  \hat{\uv}'_{k_x} (y,z,t)\,e^{\i k_x x} \right )\ .
\end{equation}
The second order cumulant variables are the ensemble mean covariances of the velocity components
of  Fourier component $k_x$  between point
$1\equiv(y_1,z_1)$ and point $2\equiv (y_2,z_2)$  evaluated at the same time:
\begin{equation}
C_{\alpha \beta,k_x}(1,2)= \left < \hat{u}_{\alpha,k_x}'(1) \hat{u}_{\beta,k_x}'^*(2)  \right >_E~,
\label{eq:covar}
\end{equation}
which is a function of  the coordinates of the two points $(1)$ and $(2)$ on the $(y,z)$ plane and of time ($*$ denotes complex conjugation). 
The SSD   equations  corresponding
to the second-order closure of \eqref{eq:RNL} are
obtained by identifying the Reynolds stress  forcing term $ \langle  \mathbf{u}' \cdot \nabla \mathbf{u}'  \rangle_x$
in  \eqref{eq:NSm}  with its  ensemble mean, $ \langle \langle \mathbf{u}' \cdot \nabla \mathbf{u}' \rangle_E \rangle_x$, 
with the fluctuations taken  from the same mean $\Uv$.
The equation for the second order cumulant can be  obtained 
by time differentiating
the covariance \eqref{eq:covar} and using  \eqref{eq:RNLp} (for a derivation cf. \cite{Srinivasan-Young-2012,Constantinou-etal-2016,Markeviciute-Kerswell_2023}).
The SSD equations in this second order closure are:   
\begin{subequations}
\label{eq:S3T}
\begin{gather}
\partial_t U_\alpha+ U_\beta \partial_\beta U_\alpha + \partial_\alpha P - \Delta U_\alpha/R = -
\frac{1}{2}  \sum_{k_x} \real \left ( \partial_{\beta,1} C_{\alpha \beta,k_x} (1,2) \right )_{1=2} \ ,
\label{eq:S3Tm}\\
\partial_t C_{\alpha \beta,k_x}(1,2)= A_{\alpha \gamma,k_x}(1) C_{\gamma  \beta, k_x}(1,2)  + A_{ \beta \gamma,k_x}^*(2) C_{\alpha \gamma,k_x}(1,2) \, \label{eq:S3Tp}
\\ 
\partial_a U_a =0 ~~,~~\hat{\partial}_\alpha (1) C_{\alpha \beta,k_x} (1,2)= \hat{\partial}_\beta^*(2) C_{\alpha \beta,k_x} (1,2) = 0\ . \label{eq:S3Tdiv}
\end{gather}\label{eq:QLE0f}\end{subequations}
with summation convention on repeated indices and  $\hat{\partial} \equiv ( i k_x , \partial_y, \partial_z )$. In \eqref{eq:S3Tp} $A_{\alpha \beta,k_x} (1)$ (or $A_{\alpha \beta,k_x} (2)$)  is the operator governing the  linear  evolution of streamwise  varying perturbations in \eqref{eq:NSp} 
with streamwise wavenumber $k_x=2 \upi n/L_x$, $n=1,2,\cdots$,
linearized about the instantaneous streamwise mean flow $\U(1)$ (or $\U(2)$) and $1$ (or $2$) indicates that the operator acts on the $1$ (or the $2$)
variable of $C(1,2)$. The subscript $1=2$ in \eqref{eq:S3Tm} indicates that after  differentiation of $C_{\alpha \beta,k_x} (1,2)$ 
with respect to  variable 1, the expression is evaluated at the same point. 
These equations produce at post-transitional  Reynolds numbers a transient chaotic trajectory of
the S3T SSD state  $(U_\alpha,C_{\alpha \beta, k_x})$,  in which only a small number of streamwise-varying components are sustained corresponding to a small set of wavenumbers, $n$.
The time-mean of the first two cumulants evolving on the S3T SSD chaotic trajectory are denoted $\langle \U \rangle_t$ and  $\langle C \rangle_t$.
This  time-mean statistical state comprises a  mean flow $\langle \U \rangle_t= (\overline{U}(y), \overline V, \overline W) $, with $\overline V= \overline W = 0$,  which   depends only on the cross-stream 
coordinate, and a second order cumulant $\langle C \rangle_t$, the components of which are 
homogeneous in the spanwise direction. 
These cumulant components are also  homogeneous in the streamwise direction, as  accounted for
by allowing the fields to be represented by Fourier decomposition in the streamwise direction.

\label{sec:framework}
\begin{figure*}
	\centering
       \includegraphics[width = .4\textwidth]{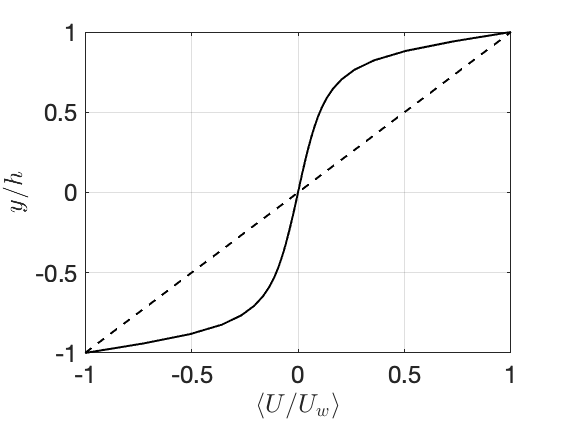}
        \caption{ The mean turbulent velocity profile, $\overline U$,   in a turbulent RNL simulation 
        of pCf at $R=600$ in a channel with $L_x/h=1.75 \upi$ and $L_z/h=1.2 \upi$. The shear at the walls is $4.2~U_w/h$.
        The laminar Couette flow is shown  for comparison (dashed).}
\label{fig:meanU}
 \end{figure*}

%
 From \eqref{eq:S3Tp} we obtain that the time mean state,  $(\langle \U \rangle_t, \langle C \rangle_t)$ satisfies the equation 
 \begin{equation}
 \overline {A}_{\alpha \gamma,k_x}(1) \overline {C}_{\gamma  \beta,k_x}(1,2)  + \overline {A}_{ \beta \gamma,k_x}^{*}(2) \overline {C}_{\alpha \gamma,k_x}(1,2) =-
 \langle \tilde A_{\alpha \gamma,k_x}(1) \tilde C_{\gamma  \beta,k_x}(1,2) + \tilde A_{ \beta \gamma,k_x}^{*}(2) \tilde C_{\alpha \gamma,k_x}(1,2)   \rangle_t~,
\label{eq:Ct}
\end{equation}
 where   $\overline C$ is the   time-mean  second order cumulant, $\langle C \rangle_t$,  $\overline A$ is the  operator governing the
 linear evolution of fluctuations on the time-mean  flow $\overline{U}(y) \hat \xv \equiv \langle \U \rangle_t$, and  tilde denotes the departures from the time mean.
 We find that the r.h.s. of \eqref{eq:Ct}
 does not vanish and consequently  $(\langle \U \rangle_t, \langle C \rangle_t)$   is not a fixed point of the S3T SSD dynamics in  post-transitional
  pCf turbulence. 
 This time-mean state would have been a fixed point of the S3T SSD dynamics if the S3T SSD dynamics had a fixed point attractor
 (other than the laminar state), as  is often the case
 in planetary turbulence. Instead, S3T dynamics demonstrates that this time-mean flow is unstable.  Identification of  this instability  ensures that the 
 ensemble statistical state  cannot be associated with a stable flow to which it 
 corresponds, as in the case of beta plane or pre-transitional boundary layer time mean states.
%
%

 We have shown that the ensemble/time mean flow is not sustained by its consistent Reynolds stresses.  
 However, it is a common practice to conjecture the existence of exogenous forces sustaining mean flows.
 Consistent with this common practice, we can  study the 
 S3T SSD stability properties of   $(\langle \U \rangle_t, \langle C \rangle_t)$ by
assuming that time-mean stresses sustain these states as equilibria
and then determine the stability of these equilibria. 
This procedure might be realized physically by adding appropriate eddy viscosity to sustain 
the time-mean velocity  as an equilibrium for the sake of studying the hydrodynamic stability properties of a flow, 
as has been done in the case of the Reynolds-Tiederman turbulent
profile \cite{Reynolds-Tiederman-1967}. 
We enforce that  $(\langle \U \rangle_t, \langle C \rangle_t)$ form an equilibrium state by introducing stresses   
calculated 
from the simulation so that the  SSD equilibrium conditions are satisfied:
\begin{subequations}
\label{eq:S3Teq}
\begin{gather}
  \Delta \overline U_\alpha/R +\overline F_\alpha=0~, \, \label{eq:Ubal}\\
 \overline A_{\alpha \gamma,k_x}(1) \overline{C}_{\gamma  \beta,k_x}(1,2)  + \overline A_{ \beta \gamma,k_x}^{*}(2) \overline{C}_{\alpha \gamma,k_x}(1,2) + \overline{Q}_{\alpha \beta,k_x}(1,2)=0~,
\label{eq:LA}
  \end{gather}
  \end{subequations} 
 with both
 \begin{subequations}
\label{eq:Feq}
\begin{gather}
\overline{F}_\alpha =-\frac{1}{2}  \sum_{k_x} \real \left ( \partial_{\beta,1} \overline{C}_{\alpha \beta,k_x} (1,2) \right )_{1=2}
-\langle \tilde U_\beta \partial_\beta \tilde U_\alpha \rangle_t +\partial_\alpha \langle P \rangle_t ~,\\
\overline{Q}_{\alpha \beta,k_x} (1,2) =  \langle \tilde A_{\alpha \gamma,k_x}(1) \tilde C_{\gamma  \beta,k_x}(1,2) + \tilde A_{ \beta \gamma,k_x}^{*}(2) \tilde C_{\alpha \gamma,k_x}(1,2)   \rangle_t ~.
\label{eq:Qinf} 
\end{gather}
\end{subequations}
obtained from the simulations.


\label{sec:framework}
\begin{figure*}
	\centering
       \includegraphics[width = 0.75\textwidth]{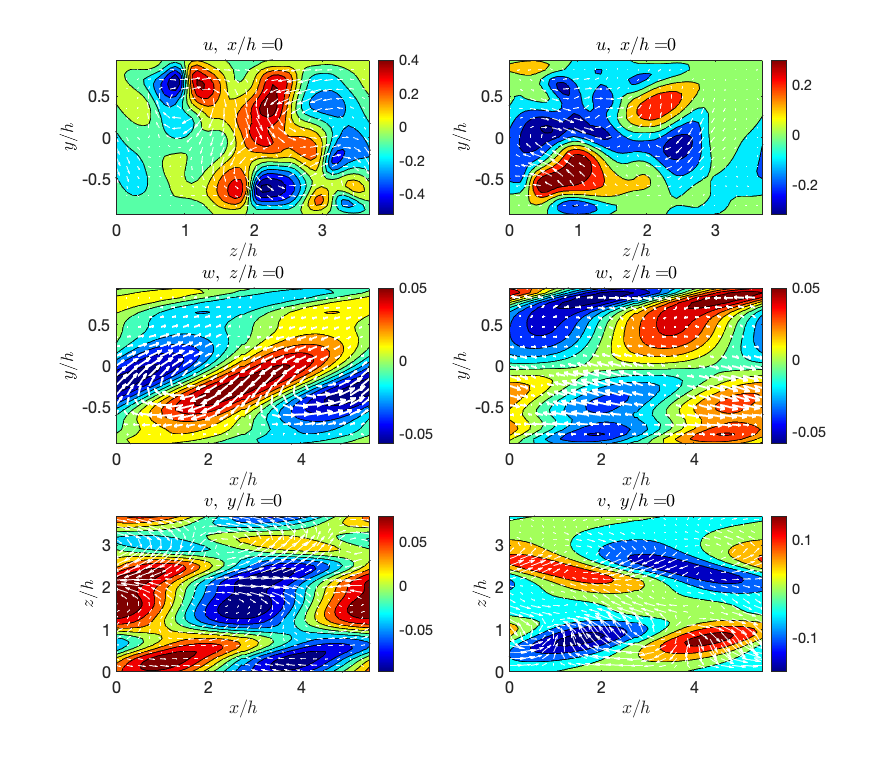}
               \caption{  Columns showing snapshots of the velocity field with $k_x=2 \pi h/L_x$ for fluctuations of the turbulent flow
               in an  RNL simulation of pCf at $R=600$. 
                The contour levels indicate the value of the fields. 
               Top row: contours of $u'$ velocity and vectors of $(w',v')$ on the $x/h=0$ plane. Maxima of  $(u', v', w')$  are  $(0.46,0.13,0.08)$ (left panel) and
                $(0.40,0.21,0.12)$  (right panel). Middle row: contours of $w'$ velocity and vectors of $(u',v')$ on the $z/h=0$ plane.  Maxima of  $(u', v', w')$  are  $(0.28,0.06,0.10)$ (left panel) and
                $(0.20,0.05,0.04)$  (right panel), Bottom row: contours of $v'$ velocity and vectors of $(u',w')$ on the centerplane, $y/h=0$. Maxima of  $(u', v', w')$  are  $(0.34,0.22,0.10)$ 
                (left panel)    and $(0.27,0.24,0.17)$  (right panel). Note that while the  Reynolds stresses  
                arising from averaging over these states are retained in $\overline{C}$,  the  phases determining the exact structures are not retained.}
\label{fig:snaps}
 \end{figure*}

\label{sec:framework}
\begin{figure*}
	\centering
       \includegraphics[width = 0.55\textwidth]{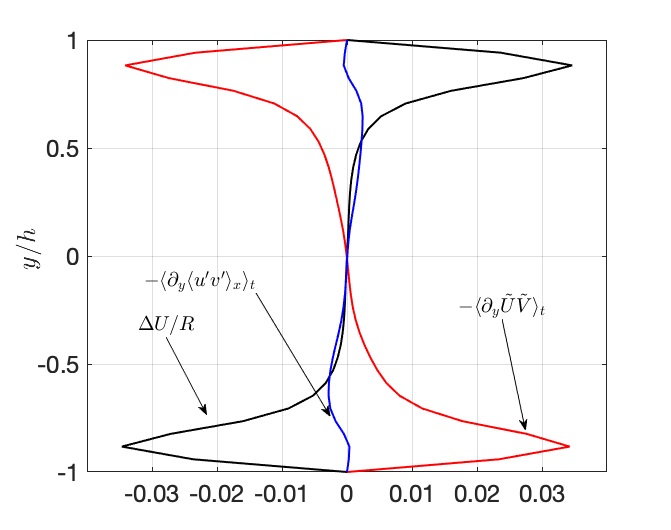}
               \caption{   The force balance of the time-mean flow in pCf at $R=600$. The viscous force
               $R^{-1} d \overline{U}/dy^2$ is primarily balanced by the divergence of the time mean Reynolds stress of the fluctuations of the streamwise-mean flow
               $-\langle \partial_y  \langle \tilde U \tilde V \rangle_x \rangle_t$. The  divergence of the  Reynolds stress of the streamwise-varying fluctuations, $-\langle \partial_y  \langle  u'  v' \rangle_x \rangle_t$  is small.}\label{fig:dennice}
 \end{figure*}

\label{sec:framework}
\begin{figure*}
	\centering
       \includegraphics[width = 0.75\textwidth]{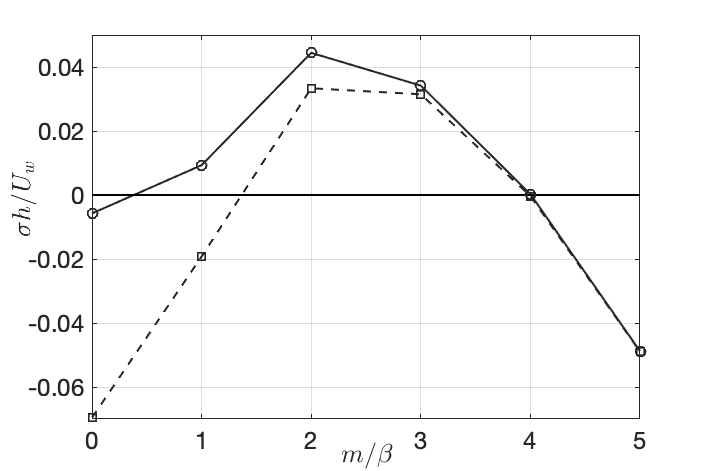}
               \caption{ Growth rates
               of the most unstable S3T eigenmodes of the turbulent time-mean  pCf mean flow shown in Fig. \ref{fig:meanU}
                as a function of the order of the spanwise harmonic, $m/\beta$, with $\beta=2 \pi h/L_z$ the fundamental spanwise wavenumber.               
                 Solid line:  growth rates of the antisymmetric in $y$ eigenfunctions. 
        Dashed line: growth rates of the symmetric in $y$ eigenfunctions.}
\label{fig:sigmaS3T}
 \end{figure*}

Having obtained in this way an equilibrium of the SSD we can determine its stability by considering the linear evolution of perturbations
$\delta U$ and $\delta  C_{k_x}$ in \eqref{eq:S3T} (cf. \cite{Farrell-Ioannou-2012,Markeviciute-Kerswell_2023}).
The perturbation equations that govern the $(\delta U,\delta  C_{k_x})$ about the time-mean state $(\overline U, \overline C_{k_x})$ are: 

\begin{subequations}
\label{eq:PS3T}
\begin{gather}
\partial_t \delta U_\alpha =  - \overline{U}_\beta ~\partial_\beta \delta U_\alpha  - \delta U _\beta ~\partial_\beta  \overline{U}_\alpha +\Delta \delta U_\alpha/R - \partial_\alpha \delta P  -
\frac{1}{2}  \sum_{k_x} \real \left ( \partial_{\beta,1} \delta C_{\alpha \beta,k_x} (1,2) \right )_{1=2} \ ,
\label{eq:PS3Tm}\\
 \partial_t \delta C_{\alpha \beta,k_x}(1,2) =  \overline{A}_{\alpha \gamma,k_x}(1) ~\delta C_{\gamma  \beta, k_x}(1,2)  + \overline{A}_{ \beta \gamma,k_x}^*(2) ~\delta C_{\alpha \gamma,k_x}(1,2)
 +\nonumber \\
  ~~~~~~~~~~~~~~~~~~~~~\underbrace{\delta {A}_{\alpha \gamma,k_x}(1) ~\overline{C}_{\gamma  \beta, k_x}(1,2)  + \delta {A}_{ \beta \gamma,k_x}^*(2) ~\overline{C}_{\alpha \gamma,k_x}(1,2)}_S  \,
 \label{eq:PS3Tp}\\
\partial_a \delta U_a =0 ~~,~~\hat{\partial}_\alpha (1) \delta C_{\alpha \beta,k_x} (1,2)= \hat{\partial}_\beta^*(2) \delta C_{\alpha \beta,k_x} (1,2) = 0\ . \label{eq:PS3Tdiv}
\end{gather}
\end{subequations}
The first equation \eqref{eq:PS3Tm} governs the evolution of the perturbation of the first cumulant, while \eqref{eq:PS3Tp} governs those of the second cumulant.
The operator  $\overline{A}_{k_x}$ is the operator that governs the evolution of perturbations about the time-mean flow $\overline{U}$,
and $\delta {A}_{k_x}$ denotes the operator that governs the evolution of perturbations about the perturbed streamwise-mean flow $\delta {U}$.
The term S in \eqref{eq:PS3Tp}, which describes the linear interaction between  perturbations to the  mean flow and the second cumulant of the unperturbed S3T equilibrium state,
is responsible for the emergence of the new S3T instabilities. When  $\overline C_{k_x}=0$
the first and second cumulant perturbation equations decouple and  linear hydrodynamic stability of  $\overline{U}$, which is determined from eigenalysis $\overline{A}_{k_x}$ in
\eqref{eq:PS3Tp}, implies the  linear S3T stability of the S3T state $(\overline{U},0)$. 
In general the $k_x$ in \eqref{eq:PS3T} are found to span only the small number of streamwise-vanumbers that comprise the active subspace sustaining the fluctuations and for which $\overline C_{k_x} \ne 0$.
In the case discussed here the turbulent state is sustained with the single streamwise wavenumber  $k_x= 2 \upi/L_x$.  For the other $k_x$ that have $\overline{C}_{k_x}=0$, no investigation is necessary as  S3T stability is implied from
the hydrodynamic stability of the flow.

\label{sec:framework}
\begin{figure*}
	\centering
	 \includegraphics[width = 0.75\textwidth]{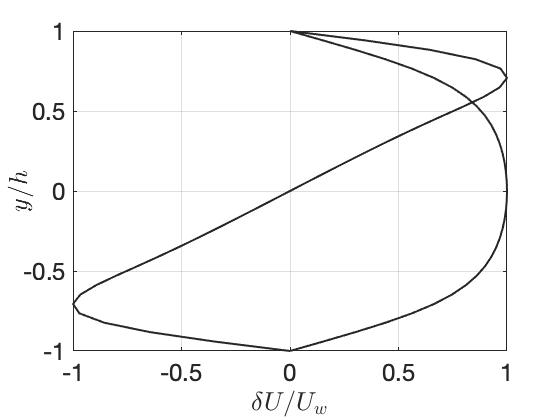}
        \caption{ {The structure of the $\delta U$ component of the first two least stable  S3T spanwise constant ($m=0$) eigenmodes of the turbulent pCf, for which $\delta V=\delta W=0$.
        The time-mean flow when maintained to be a S3T equilibrium is stable to $m=0$ (spanwise constant)  but  unstable to spanwise varying perturbations, as shown in Fig. \ref{fig:sigmaS3T}. 
          The antisymmetric perturbation is the least stable with decay rate $\sigma =-0.006 ~U_w/h$ while the least stable  symmetric perturbation has
        decay rate $\sigma=- 0.07~U_w/h$.  These stable $m=0$ modes do not correspond to the least stable eigenfunctions of the dynamics of relaxation to the time-mean flow, which are  shown in Fig. \ref{fig:Amodes}.}}	
\label{fig:modeS3Tm0}
 \end{figure*}

\label{sec:framework}
\begin{figure*}
	\centering
	 \includegraphics[width = 0.75\textwidth]{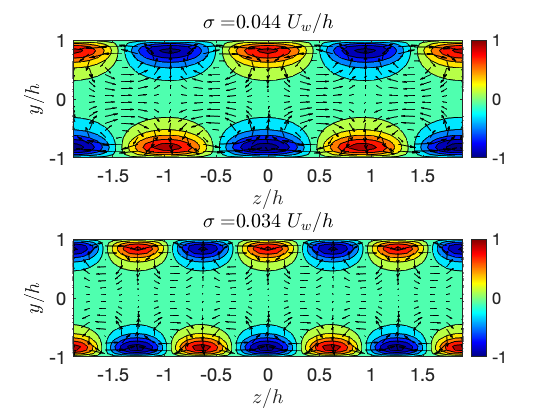}
        \caption{ The structure of the first two most unstable eigenmodes of the turbulent pCf mean flow shown in Fig. \ref{fig:meanU} 
        obtained by eigenanalysis of  the operator governing the evolution of perturbations in the S3T equations. The most unstable eigenmode with 
        spanwise wavenumber
        $m = 2 \beta$ (top panel)  has growth rate $\sigma=0.04~U_w/h$, while the second most unstable eigenmode 
        with $m= 3 \beta $ (bottom panel) has growth rate $\sigma=0.034~U_w/h$. The first cumulant component of the S3T eigenfunction
       has   roll-streak structure. Shown are contours of the streamwise mean velocity, $\delta U$ 
       (the contour level spacing is $0.2$),  and velocity vectors of the components $(\delta V, \delta W)$  on the $(y/h,z/h)$ plane. 
       Consistent with the lift-up mechanism, positive $\delta V$ is associated with negative $\delta U$. 
       The maxima of $(\delta U, \delta V, \delta W)$  are proportional to $(1,0.23,0.26)$ in the $m=2\beta$ S3T eigenfunction
       and proportional to $(1,0.25,0.22)$ in  the  $m=3\beta$ S3T eigenfunction.}       
\label{fig:modeS3T}
 \end{figure*}

\section{Results}

We consider a turbulent pCf at $R=600$ in a channel with $L_x/h=1.75 \pi$, $L_z/h=1.2 \pi$ 
in the quasi-linear approximation i.e. neglecting the fluctuation-fluctuation interaction in \eqref{eq:NS},
as discussed in the previous section. The turbulent state  supports fluctuations of only the single $k_x=2 \pi h /L_x$ wavenumber.

We obtain the flow states for a period of $10^4 ~h/U_w$ time units  with a discretization on $N_y=33$ grid points in $y$ and $N_z=48$ grid points in $z$.  
We have verified that the time period of the simulation is adequate for producing  converged results \footnote{Convergence to the
homogeneous statistical symmetry  in wall-bounded  flows is very slow. The convergence towards statistical symmetry in pPf  is
shown in the Appendix of \citet{Nikolaidis-POD-2023}.}.
{\color{black} The time-mean turbulent flow $\overline U (y)$ is  shown in Fig.  \ref{fig:meanU}, and 
the  structure of two typical fluctuation states are shown in Fig. \ref{fig:snaps}.  $\overline{C}$ is the time-mean  covariance of the velocities of the streamwise-varying flow 
obtained from  all the individual snapshots that occur
in the simulation. In the process of forming  $\overline{C}$,  the phase information associated with  individual fluctuations is lost, while the  time-mean Reynolds stresses
that balance the time-mean flow are retained.  Further insight into the maintenance of the time-mean flow can be obtained
from the individual terms contributing to the maintenance. The time-mean flow force balance \eqref{eq:Ubal} for the case of Couette 
turbulence simplifies to:
\begin{equation}
\label{eq:2D3C}
  \underbrace{\frac{1}{R} \frac{d^2 \overline{U}}{dy^2}}_A - \underbrace{\frac{d \langle \langle  {u' v'} \rangle_x \rangle_t}{d y}}_B - \underbrace{\frac{d \langle  {\tilde U \tilde V} \rangle_t }{d y}}_C=0~,
  \end{equation}
with the viscous force (term A), by which the boundaries maintain the flow,   being almost exclusively   balanced by 
the time-mean Reynolds stress divergence of the  streamwise constant time-varying velocity components (term C) (cf. Fig. \ref{fig:dennice}),
consistent with the mechanism of streak displacement by roll circulations in the 2D3C turbulence model of 
\citet{Gayme-etal-2010}.
 It is remarkable that the component of the Reynolds stress divergence arising from the streamwise constant flow  fluctuations  (term C) 
so dominates the mean velocity force balance. This calls into question the 
adoption of spatially dependent diffusion as a mechanistic explanation for momentum fluxes maintaining time mean flows insofar as these fluxes are predicated on 
assuming
they arise from small scale velocity correlations.
The stress divergence arising from the 
time-mean eddy Reynolds stress of the streamwise varying components, 
obtained from the covariance $\overline{C}$  (term B), makes a minor  contribution to the force balance \eqref{eq:2D3C}.}

{\color{black} While $\overline C$  is by necessity positive definite, being the time average of the instantaneous positive definite covariances of the 
fluctuations, this is not guaranteed for the time-mean $\overline{Q}$ obtained in \eqref{eq:Qinf}.
 We find that $\overline{Q}$
 is non-positive definite, as was also found by \citet{Zare-etal-2017} in  a  DNS of wall-bounded turbulence.
 This implies that if $\overline{Q}$ were to represent  the spatial covariance 
 of a stochastic excitation, this stochastic excitation must be temporally correlated (colored) \cite{Zare-etal-2017}.}

{\color{black} The operator  $\overline A_{k_x}$ in \eqref{eq:LA} governing the evolution of perturbations about the time--mean flow $\langle \U \rangle_t \equiv \overline{U}(y)$
is stable with the least damped mode  at  streamwise wavenumber  $k_x=2 \pi h/L_x$ decaying  rate  $\sigma=-0.11 ~U_w/h$.}  
As is the case for the Reynolds-Tiederman mean-flow, the time-mean  turbulent pCf flow, $\overline{U}(y)$, is far from a 
state of marginal hydrodynamic stability, 
violating the conjecture of Malkus that the time-mean flow is adjusted by turbulence to neutral stability, as occurs in 
Rayleigh-B\'enard convection \cite{Malkus-1956,Reynolds-Tiederman-1967}.
However, here we show that the time-mean  flow,  although hydrodynamically stable,
is unstable to the nonlinear instability  revealed by the S3T dynamics.

To proceed with the S3T stability analysis of the turbulent state  $(\langle \U \rangle_t, \langle C \rangle_t)$ 
we consider the stability of   equations  \eqref{eq:PS3T} governing cumulant perturbations  about $(\langle \U \rangle_t, \langle C \rangle_t)$. The discretized linear operator that 
governs perturbations $(\delta U, \delta V,\delta W,\delta C_{k_x})$ in \eqref{eq:PS3T} 
has dimensionality $2N_yN_z (2N_yN_z +1) \times2N_yN_z (2N_yN_z +1)$, which is   $\approx 10^7 \times 10^7$ in the model example 
we study, necessitating the use of 
 the power method
 to obtain the fastest growing eigenvalues and eigenmodes of the  equilibrium state,  $(\langle \U \rangle_t, \langle C \rangle_t)$,
 as discussed in \cite{Farrell-Ioannou-2012,Farrell-Ioannou-2017-bifur}. 
  Due to the spanwise homogeneity of the background state 
 the perturbation eigenmodes are harmonic  in $z$  with mean flow components,
 $(\delta U_m(y),\delta V_m(y),\delta W_m(y))e^{i mz} $,
 where  $m$ is the spanwise wavenumber, 
 with
 corresponding covariance $\delta C_{m,k_x} (y_1,y_2) e^{i m (z_1-z_2)}$.
  The growth rates of the most unstable S3T symmetric and antisymmetric  in $y$ eigenfunctions as a function of $m$
 are shown in Fig. \ref{fig:sigmaS3T}.  While the spanwise constant perturbations $(m=0)$ are S3T stable with structure shown in Fig. \ref{fig:modeS3Tm0}, 
 the introduction of spanwise variation  results in  the time-mean flow  being unstable to  $m=1\beta,2\beta,3\beta$  S3T perturbations, with $\beta=2 \pi h/L_z$ the
 fundamental spanwise wavenumber. 
The  $(\delta U_m(y),\delta V_m(y),\delta W_m(y))e^{i mz} $,
  component of  the two most unstable eigenmodes of the perturbation S3T dynamics, with $m=2 \beta$ and $m=3 \beta$, are shown 
in Fig. \ref{fig:modeS3T}.  These unstable  eigenmodes have the  form of streamwise roll-streak (R-S) structures.
This is also the case for the 
S3T eigenmodes of the laminar profile,
but in contrast to the laminar flow S3T eigenmodes these  R-S structures are confined to the near wall high shear regions of the mean turbulent
profile. This universality of the structure of the S3T eigenfunctions reflects the universality in the mechanism of the S3T R-S instability
\cite{Farrell-Ioannou-2012,Farrell-Ioannou-2022}. The two most unstable eigenmodes have real eigenvalues with growth rates  $\sigma= 0.04 ~U_w/h$ and 
$\sigma= 0.034~U_w/h$ for spanwise wavenumber 
$m= 2 \beta$ and $m=3 \beta$ respectively. Both eigenmodes are antisymmetric in $y$. 
Their symmetric counterpart eigenmodes are also unstable but with the smaller growth rates $\sigma= 0.025 ~U_w/h$ and $\sigma= 0.032~U_w/h$  respectively.
As shown in Fig. \ref{fig:sigmaS3T}, degeneracy of the growth rates between the symmetric and antisymmetric in $y$  modes is approached
as $m$ increases.  Degeneracy  is expected 
as the velocity fields associated with the modes become more confined to their respective  boundaries  with either increase in $m$ or $R$.

We conclude that the S3T-SSD equilibrium state in pCf is unstable to perturbations 
with R-S structure and therefore can not be used to obtain the linear dynamics of 
perturbations from the mean statistical state by performing a perturbation analysis of the S3T SSD about the  mean 
turbulent state. We note that the  R-S unstable modes in turbulent flows depend 
only on the shear and similar unstable structures are 
expected to be found in other wall-bounded flows such as pPf and boundary layers.
{\color{black} The fact that an ensemble mean state can be stable in ensemble dynamics while the same state is unstable to 
perturbation of each ensemble member individually  can be understood by recalling the example of a mass-spring with random restoring
force governed by the Mathieu equation \cite{Farrell-Ioannou-2002-perturbation-I}.
Each realization of the state is unstable
while the ensemble mean evolves as a stable harmonic oscillator with the average square
frequency of the ensemble realizations. This example shows how the time-mean state may
be an attractor for the ensemble mean but at the same time be a repeller for the perturbation
dynamics as is the case for wall-turbulence.}

\section{Empirical determination of the effective linear  dynamics of perturbations to the time-mean equilibrium state in a turbulent channel flow}

Perturbation analysis of the S3T SSD equilibrium state provides comprehensive characterization of 
turbulent state stability in cases for which a stable S3T SSD equilibrium state exists.  
However,   turbulent  pCf and pPf lie on a transient chaotic attractor 
in statistical state  space.  Stability of the ensemble mean (or equivalently the time-mean) statistical state 
in these cases requires obtaining the perturbation dynamics
averaged over the chaotic attractor. This can be accomplished by restricting  attention to the 
time-mean flow component of the statistical state. 
As discussed in the introduction,  IWCV  found  these first two eigenmodes and eigenvalues  by averaging  an ensemble of
pPf  DNS runs at $R_\tau=180$ in  a channel with  $L_x/h=8 \pi $ and $L_z/h=4 \pi$, $L_y/h=2$,
which had been perturbed by the same initial streamwise flow perturbation, $\delta U(y)$. 
Here we 
obtain an approximate  dynamical system in Langevin form for the relaxation of mean flow perturbations to the time-mean
using an alternative method, called linear inverse modeling.  The linear inverse model (LIM) obtains  the 
effective dynamics of perturbations
to the time-mean  flow by observing the behavior of 
fluctuations to the time-mean velocity
profile naturally occurring in the turbulence.
{\color{black}In our study we first obtain using LIM the empirical  dynamical system that governs 
the fluctuations from the streamwise-spanwise mean flow  
in the pCf simulation presented in the previous section, sampled every non-dimensional time unit, which we verified to adequately sample the temporal  fluctuations of the streamwise 
and spanwise mean flow and also in a 
pPf at $R_\tau=180$ in a channel of $L_x/h=2\pi$ and $L_z/h=\pi$, $L_y/h=2$, with bulk velocity, $U_b$.
The pPf data were  obtained using   a constant mass-flux  DNS. The pPf and the pCf were integrated over a sufficient time ($6 \times 10^4 ~h/U_b$ for the pPf and $10^4 ~h/U_w$ for the pCf)
to obtain converged results as verified by using the data for half the interval.}
 \label{sec:framework}
\begin{figure*}
	 \includegraphics[width = 0.75\textwidth]{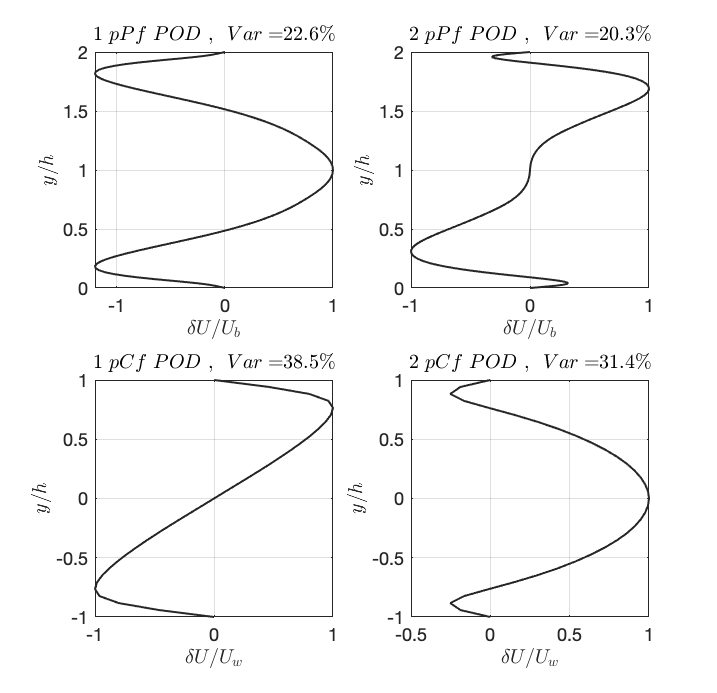}  
        \caption{ The structure of the first two POD modes of the mean flow fluctuations in  pPf (top), and in pCf (bottom)
         Top left panel:  the first POD mode of pPf accounts for $23\%$ of the energy of the fluctuations. Top right panel:
         the second POD of pCf accounts for $20\%$ of the energy of the fluctuations.
         Bottom left panel:  the first POD mode of pCf accounts for $39\%$ of the energy of the fluctuations. Bottom right panel:
         the second POD of pCf accounts for $31\%$ of the energy of the fluctuations.    }
\label{fig:PODmodes}
 \end{figure*}
%
\label{sec:framework}
\begin{figure*}
       \includegraphics[width = 0.75\textwidth]{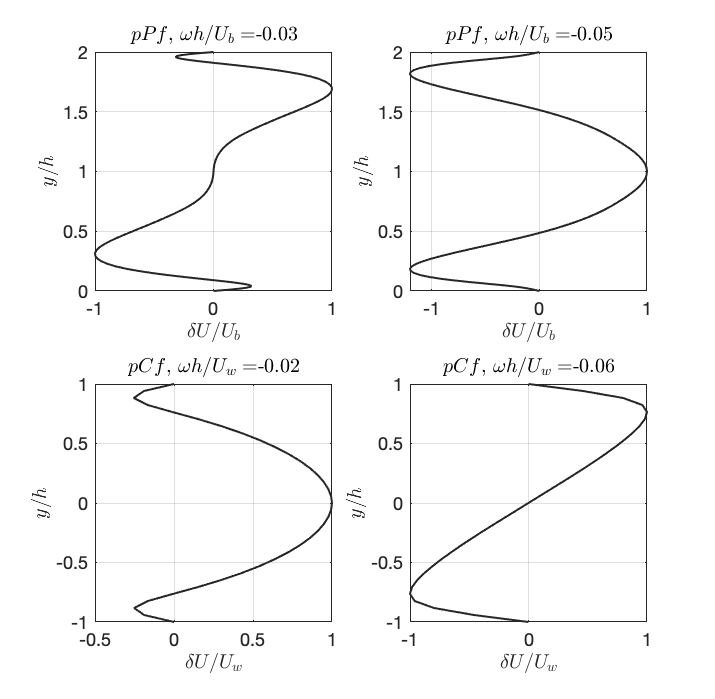}	
        \caption{ Structure  of the     eigenmodes of the
        $2\times2$ LIM operator governing the relaxation of steamwise and spanwise mean  flow fluctuations to the time-mean flow. Top left panel: the least damped mode in pPf is antisymmetric about the channel center
       and  has eigenvalue $\omega h/U_b= -0.03$. Top right panel: the second least damped mode  is symmetric and has
        eigenvalue  $\omega h/U_b=-0.05$.  Bottom left panel: the  least damped mode in pCf  is symmetric about the channel center
        and have eigenvalues $\omega h/U_w= -0.02~$. Bottom right panel:  the second least damped mode is antisymmetric about the channel center
        and have eigenvalues $\omega h/U_b= -0.06$.}
\label{fig:Amodes}
 \end{figure*}

From the data we obtain the time series of the fluctuations $\delta U(y,t)$ about the time-mean flow
{\color{black} with the asymptotic statistical symmetry of the time-mean flow, $\overline{U}(y)$, about the center-plane in the cross-flow direction
in pPf  and antisymmetry in pCf  enforced  by adding  replicas  of the mirror symmetric instantaneous mean flows about the center-plane in pPf
and mirror antisymmetric replicas in pCf. Although the instantaneous realizations of the streamwise-spanwise averaged flow are not symmetrized, 
this symmetrizing operation results in the  time-mean flow obtained from our finite data set, as well as in the time-mean statistical 
moments of the fluctuations to be 
exactly symmetric in pPf and antisymmetric in pCf. An important consequence of this symmetrization procedure
is that it enforces in the data the known symmetries of the dynamics of pCf and pPf. }

From the $\delta U(y,t)$ fluctuations we also obtain the time-mean  fluctuation covariance:
\begin{equation}
C_0(1,2)= \langle \delta U(y_1,t) \delta U(y_2,t) \rangle_t ,
\label{eq:C0}
\end{equation}  
and the time-mean time-lagged covariances:
\begin{equation}
C_\tau(1,2)= \langle \delta U(y_1,t+\tau) \delta U(y_2,t) \rangle_t.
\label{eq:Ctau0}
\end{equation}
{\color{black} 
Due to the mean flow symmetrization procedure, the covariance  $C_0$ and $C_\tau$
are  symmetric about the $x-z$ plane at the  channel center, i.e. they satisfy
$C(y_1,y_2)=C(y_1^s,y_2^s)$,
where $y_i^s$, $i=1,2$, is the symmetric  coordinate of $y_i$ with respect to the channel center.}  
The POD modes of the mean flow fluctuations are obtained from eigenanalysis of $C_0$.
Due to the statistical symmetry reflected in $C_0$, the POD modes are necessarily  either symmetric or antisymmetric.
%
 
LIM determines the best fit to our data by the linear stochastic system with Langevin form:
\begin{equation}
\frac{d ~\delta U}{dt} = A ~\delta U +  \xi(t)~,
\label{eq:Langevin}
\end{equation}
where $\delta U$ is the column vector  of the values at the $N_y$ wall-normal grid points,  $A$ is the   
$N_y \times N_y$ matrix generator of the dynamics of  fluctuations
and $\xi(t)$ represents the spatial and temporal 
structure of the unresolved  dynamics that are  required to be parameterized as 
a zero mean temporally delta correlated Gaussian noise process  satisfying $\langle \xi(y_1,t) \xi(y_2,t')  \rangle_E = \Xi(1,2) \delta(t-t')$,  
with the  full rank  positive definite spatial covariance $\Xi(1,2)$ 
to be determined, along with the effective linear operator $A$,
 by inversion for the best fit to the dynamics \eqref{eq:Langevin}.
 {\color{black}  Note that LIM does not determine $\xi$, it determines only  its spatial covariance $\Xi$.}
{\color{black} In  LIM  modeling the operator $A$ 
is discovered and is not assumed a priori  to be either  the operator $\overline{A}$ that governs perturbations about the mean flow, $\overline{U}$, or $\overline A$ modified by eddy viscosity
as in the studies of \citet{Zare-etal-2017,Hwang-Eck-2020,Towne-Lozano-2020,Holford-etal-2023,Holford-Hwang-2023,Abootorabi-Zare-2023}.}

 With a discretized representation  the  time lagged covariances of the velocity fluctuations, $C_0$ and $C_\tau$ in \eqref{eq:C0},\eqref{eq:Ctau0}, become  matrices. 
Under the assumption that the stochastic excitation, $\xi$, in  \eqref{eq:Langevin}  is temporally white it can be shown (cf. \cite{DelSole-04}) that:
 \begin{equation}
 C_\tau =e^{A \tau} C_0~, \tau > 0~,
 \label{eq:Ctau}
 \end{equation}
 from which we obtain immediately that: 
 \begin{equation}
 A = \frac{1}{\tau} \log (  C_\tau C_0^{-1} )~,
 \label{eq:A}
  \end{equation}
{\color{black}The  logarithm of any matrix $M$ that has a full basis of eigenvectors  is 
$\log(M)=U D U^{-1}$, where $U$ is the matrix of the eigenvectors of $M$ and $D$
the diagonal matrix with diagonal elements
the logarithms of the eigenvalues of $M$. 

If there is an interval in  $\tau$ for which,   $\tau$ is both small enough to sample the dynamics and large enough 
that the correlation time of the fluctuations perturbing the eigenmodes can be 
represented by a Gaussian delta-correlated white noise process,  so that  in this interval 
$A$ is  insensitive to  $\tau$, then $A$ identifies the  generator of the Langevin dynamics  \eqref{eq:Langevin}
rather than being merely the finite time map connecting $C_0$ to  $C_\tau$
\cite{Papanicolaou-Kohler-1974,Penland-1995a,DelSole-2000,Penland-2019}.}

 Boundedness of $C_0$ ensures  the stability of the matrix $A$   and 
 \eqref{eq:Langevin} 
 requires that the covariance of the noise process $\Xi$    satisfies   the Lyapunov equation:
 \begin{equation}
 A C_0 + C_0 A^T = - \Xi~.
 \label{eq:Lyap}
 \end{equation}
{\color{black} $\Xi$ must be verified, for consistency,  to be  positive semi-definite
given that it is the correlation of  the forcing vector structures
 \cite{Penland-1995a}.
In this way LIM  identifies the linear operator, $A$,  that best fits the data while also determining the  positive semi-definite spatial covariance, $\Xi$, 
of the effectively  white  temporal excitation of the fluctuations. 
By contrast, note that \citet{Zare-etal-2017} constrain their stochastic model operator to 
be $\overline{A}$, which governs perturbations about the time-mean flow $\overline{U}$. The consequence  is
that the $\Xi$ that  is obtained using \eqref{eq:Lyap} may not be positive definite. Such a non positive $\Xi$ is
compatible with a set of  stochastic models  
excited
by colored noise \cite{Georgiou-2002,Zare-etal-2017}.
To remove ambiguity among the members of this set \citet{Zare-etal-2017} employ an optimization procedure.
Moreover, \citet{Georgiou-2002,Zare-etal-2017a,Zare-etal-2017} show  that each of these 
colored noise models with linear operator $A_{col}$ corresponds to a  linear 
model with an appropriately modified operator $A_{white}$
driven by a white noise process. Importantly, LIM, removes the ambiguity in the choice of $A_{white}$ 
 by determining the  operator  that best fits  the data. LIM   determines a unique operator by 
using  the extra information encoded in the time lagged covariances, while \citet{Zare-etal-2017} utilize only the information encoded in $C_0$.  
A clue to the underlying importance of identifying the unique  LIM modified  operator $A$ is offered  in
 \citet{DelSole-Farrell-1996} who show   that the modified $A$ in two-layer baroclinic turbulence 
 model is the linear operator about the time mean flow, $\overline{A}$,
 corrected by the addition of  a diffusion operator. This is consistent with the works of
 \citet{Hwang-Eck-2020,Holford-etal-2023,Holford-Hwang-2023,Abootorabi-Zare-2023} who match the observed covariances by forcing white 
 the linear operator about the mean flow, $\overline A$, modified by the inclusion of
 a diffusive operator with  the Reynolds-Tiederman diffusion coefficient, indicating that the color of the stochastic forcing
 can be compensated for by inclusion of a parameterization of eddy viscosity in the form of a 
 spatially varying diffusion. However, LIM  allows going a step further by obtaining the unique form of this diffusion
 and by so doing providing an independent justification 
 for the use of eddy diffusion as a parameterization of unresolved scales in turbulence models.}

Another consideration for obtaining the  effective dynamics is that  
the inverse of $C_0$ is ill-conditioned for data from a turbulent flow
given that any finite time series can not resolve 
fluctuations of arbitrarily small scale and amplitude
(cf. \citet{North-1982}). Therefore, in order to obtain a converged dynamics  we confine the representation of the dynamics to
 the subspace spanned by a set of dominant POD modes obtained from  the covariance 
$C_0$ of the $\delta U$ fluctuations.  

We  first obtain the effective $2\times 2$ dynamical operator $A$ governing  the dynamics  of the fluctuations to the time-mean flow in the space of the  top 2 POD modes
in the pCf S3T  and the pPf DNS, which account for $\approx 70\%$  and $\approx 40\%$ of the energy respectively. 
 The top two POD modes are a symmetric and  antisymmetric pair  with 
structure   shown in Fig. \ref{fig:PODmodes}.
 We  obtain  the operator, $A$,
from \eqref{eq:A} and verify that the the noise covariance, $\Xi$, 
from \eqref{eq:Lyap} is positive definite.
Insensitivity of the  dynamics was obtained   for time-lags  around  $\tau U_b/h= 18 $ in both   pCf and pPf. 
The eigenmodes of $A$ are found to have  the same structure as the POD modes
used in the projection. 
The  eigenvalue of the least 
damped antisymmetric mode is  $\sigma=-0.03~U_b/h$
 and that of the symmetric mode is $\sigma=-0.05~U_b/h$.
 The  least damped eigenmode,  shown in the top left panel of Fig. \ref{fig:Amodes}, is antisymmetric with respect to the channel center,
consistently with the results of  IWCV.  However, our LIM analysis predicts a slower decay rate than theirs. 
The next in decay rate mode of $A$, shown in the top right panel of Fig. \ref{fig:Amodes},  is symmetric  and its decay
rate and structure are consistent with   the least damped symmetric eigenmode
of IWCV. 
The least damped eigenmodes in pCf are also real as shown in the lower panels of  Fig. \ref{fig:Amodes}. 
The eigenvalue of the symmetric mode is $\sigma= -0.02~U_w/h$, and that of the asymmetric mode is $\sigma=-0.06~U_w/h$.
 While the POD modes have the same structure as the eigenmodes of $A$, 
 in pPf the least damped antisymmetric mode of $A$ is excited less vigorously
by $\Xi$,  with the  result that  the  
dominant POD becomes the symmetric, although this is the more damped eigenmode of $A$. Similarly,
in  pCf the antisymmetric mode is favored by the excitation so that 
it accumulates more energy in the mean than does the least damped symmetric mode.

{\color{black} Coincidence of the modes of $A$ with the orthogonal  POD modes    imply  the normality of $A$.
Note that reflection  symmetry of the ensemble dynamics about the cross-stream channel center requires that  symmetric structures in the ensemble dynamics
 evolve to symmetric structures and antisymmetric to antisymmetric.
Consequently, in the linear dynamics of the ensemble  the eigenmodes of the dynamics are partitioned into a mutually orthogonal
 symmetric and antisymmetric set and the covariance of the excitation does not mix the symmetric and antisymmetric subspaces.
An example of this property is the diagonal structure of  both $A$ and of the spatial covariance,  $\Xi$, in the $2\times2$ projection of the dynamics.
  From this observation we also understand that any non-normality  which 
may arise at higher truncations will be limited to the 
 respective symmetric and antisymmetric subspaces (cf. Appendix for a discussion of the case  of a $6\times 6$ truncation). }

\label{sec:framework}
\begin{figure*}
      \includegraphics[width = 0.75\textwidth]{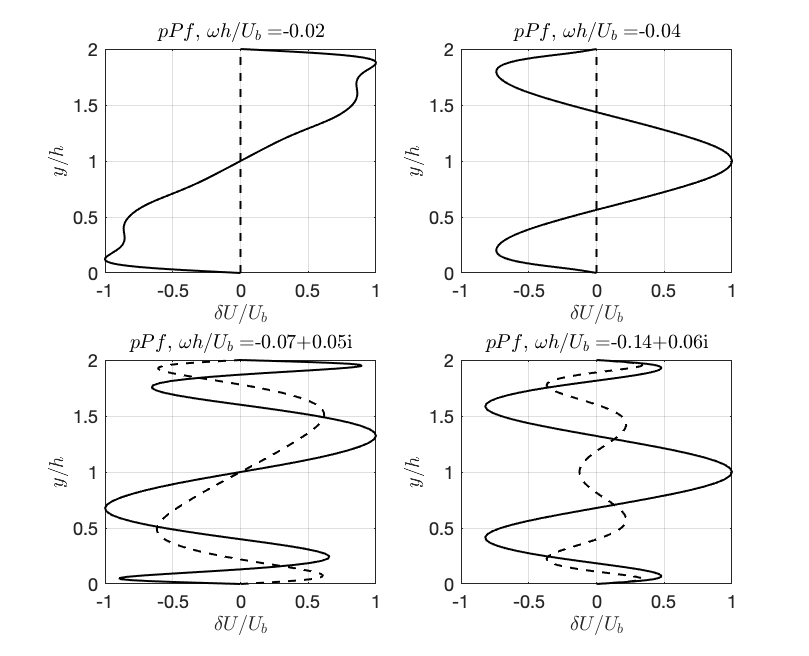}
         \caption{ The structure of the real part (solid line) and the imaginary part (dashed line) of the  6   damped eigenmodes of the
        empirical operator $A_{6}$ of the pPf. Top left panel: the least damped mode is antisymmetric about the channel center
       and  has eigenvalue $\omega h/U_b= -0.02$. Top right panel: the next least damped mode  is symmetric and has
        eigenvalue  $\omega h/U_b=-0.04$.  Bottom left panel: the third and fourth least damped modes are antisymmetric about the channel center
        and have eigenvalues $\omega h/U_b= -0.07\pm0.05 i~$. Bottom right panel:  the fifth and sixth least damped modes are symmetric about the channel center
        and have eigenvalues $\omega h/U_b= -0.14\pm0.06 i$. The structure of the complex valued eigenmodes  periodically vary with a period of $T= O(100)~h/U_b$.  }        
\label{fig:Amodes6}
 \end{figure*}

\label{sec:framework}
\begin{figure*}
       \includegraphics[width = 0.75\textwidth]{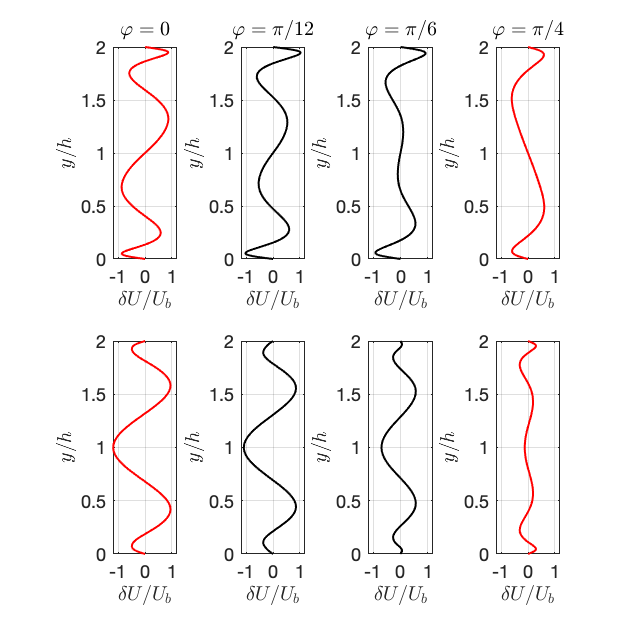}
        \caption{ Evolution of the structure of the complex eigenmodes of $A_6$ over a quarter period of their oscillation. The decay is not included in this representation.      
        Top panels: for   the antisymmetric eigenmode shown in Fig. \ref{fig:Amodes6}c  with period $125 ~h/U_b$. Bottom panels:  for the symmetric eigenmode in Fig. \ref{fig:Amodes6}d.  
        with period $105 ~h/U_b$.  }
\label{fig:vacci}
 \end{figure*}

 \label{sec:framework}
\begin{figure*}
       \includegraphics[width = 0.75\textwidth]{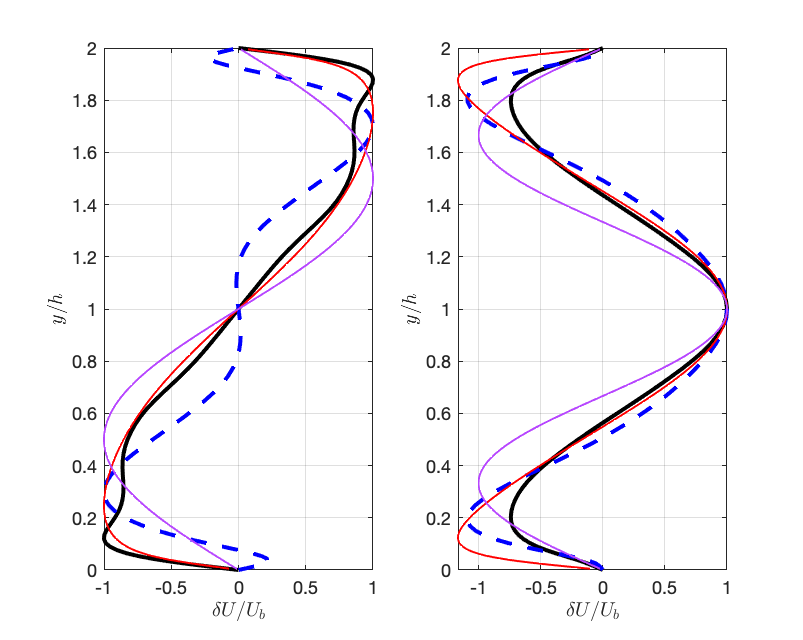}
        \caption{ Comparison of the structure of the antisymmetric (left panel) and symmetric (right panel) least damped mode
        of $A_6$ (black line) and of the corresponding modes  in diffusive  models. With red lines: the least decaying modes
       of the  Reynolds-Tiederman eddy-viscosity model at $R_\tau=180$ (red). With purple solid lines: 
       the  least damped modes  of the purely diffusive model  with $R=2767$, corresponding to the laminar Reynolds number associated with $R_\tau=180$.
       The structure of the top POD modes is indicated with the dashed blue line. The departure of the $A_6$ modes
       from the associated POD mode indicates the degree of non-normality of the ensemble dynamics of relaxation to the mean and the degree of non-commutation of $A_6$ with $\Xi_6$.}
\label{fig:eddy}
 \end{figure*}

 Given the constraints on the dynamics arising from projections onto a single symmetric and a single antisymmetric POD 
mode,
we consider higher order 
truncations.
As already discussed, non-orthogonality can only 
emerge among the  antisymmetric  (or symmetric) eigenmodes
of $A$ and  other antisymmetric (or symmetric) eigenmodes of $A$.  We demonstrate the emergence of non-normality in the dynamics in
the case of the pPf by  retaining in the dynamics  the first 6 POD modes, which together account for $90 \%$ of the mean-flow fluctuation energy.  
The results that we present do not appreciably change
when the dimension of the retained POD modes is increased to 10.
With 6 POD modes  we find that $\tau = 18 ~h/U_b$ produces  a $\tau$-insensitive  dynamical operator, $A_6$,
and converged positive definite covariance, $\Xi$ (cf. Appendix for details). 
We conclude that in pPf the dynamics of perturbations to the time mean flow is generated by the matrix $A_{6}$ the eigenvalues of which are:
$-0.02~U_b/h$, $-0.04~U_b/h$, $(-0.07 \pm 0.05 i)~U_b/h$, $(-0.14 \pm 0.06 i)~U_b/h$.
The  least damped eigenmode,  shown in the top left panel of Fig. \ref{fig:Amodes6}, is antisymmetric with respect to the channel center,
consistent with the results of  IWCV. 
However, our LIM analysis predicts a slower decay rate than theirs 
for the same $R_\tau$,  allbeit in a channel of 1/4th  size.    The next in decay rate mode of $A_6$ is 
symmetric in structure
with decay
rate and structure consistent with  the real part of the least damped symmetric eigenmode
of IWCV. {\color{black} Although the first two eigenvalues of $A_6$ are real and represent the ensemble mean relaxation of structures of fixed form  to the mean flow with time scales of $\approx 50~h/U_b$
and $\approx 25~h/U_b$ respectively,
the  remaining eigenmodes  represent a damped oscillatory relaxation  to the mean flow  which oscillates between  the real part and the imaginary part of the eigenmode
over $\approx 100 ~h/U_b$, which is  the  characteristic  period of the SSP cycle \cite{Hamilton-etal-1995}.
The evolution of the oscillating structure of the two complex eigenmodes is shown in Fig. \ref{fig:vacci}.

The operator $A_6$ governing the dynamics is   non-normal in the energy norm, as was conjectured  by   IWCV.
The inner product of the eigenmodes shows that there are substantial projections among modes in both the  symmetric and antisymmetric subspace
and therefore the operator $A_6$ is substantially non-normal (cf. Appendix).  The difference in structure between  the eigenmodes and the POD modes
reflects both the degree of non-normality of $A_6$ and the degree of non-commutation of $A_6$ with the excitation covariance 
$\Xi_6$ \cite{North-84,North-etal-2009}; in general the POD modes in this Langevin system coincide with the eigenmodes of the linear operator $A$  when $A$ is normal
and $[A,\Xi]=0$ (see Appendix).}
Fig. \ref{fig:eddy} shows that the structure of the eigenmodes departs appreciably from that of the associated POD modes. Moreover,  
non-normality of the dynamics  is not consistent with  a  strictly  diffusive process.
However, it could be argued that the flow relaxes through the smoothing that results from the advection induced by  the random action of the streamwise roll motions
over  the ensemble realizations of the flow and therefore the dynamics might be modeled as a diffusion with an appropriate coefficient
assuming diffusion to be broadly conceived as any transport process resulting in a linear flux/gradient relation.
To evaluate this hypothesis we compare the decay rates and structures of the eigenmodes predicted by an
eddy viscosity model in which the mean flow
perturbations
are governed
by
\begin{equation}
\frac{\partial  \delta U}{\partial t} =  \frac{\partial}{\partial y} \left ( \nu_E(y) \frac{\partial \delta U}{\partial y} \right )~,    
\end{equation}
with $\nu_e(y)$ an eddy viscosity coefficient.
We will obtain the eigenvalues and eigenmodes in the case of   simple diffusion  with $\nu_e(y)=1/R$, with $R$ the laminar value of the Reynolds number, and
the eddy viscosity that sustains the Reynolds \& Tiederman \cite{Reynolds-Tiederman-1967} turbulent profile
in a pPf channel with walls at $y=0$ and $y=2$.   This profile is maintained by
$
\nu_E = (1+E(y))/R_\tau
$
and  $E(y) =\left ( \sqrt{1+e^2}-1 \right)/2$, where
\begin{equation}
e(y)=\frac{k R_\tau}{3} (1-(y-1)^2) (1+2(1+(y-1)^2)(1-\exp(-R_\tau (1-|y-1|)/A))~,
\end{equation}
with $K=0.525$ and $A=37$ as is appropriate for $R_\tau=180$.  
The least damped modes comprise a symmetric/antisymmetric pair, shown 
in Fig. \ref{fig:eddy}. These modes
are qualitatively similar to the modes obtained by LIM (cf. Fig. \ref{fig:eddy}), 
but lack the detailed structure of the LIM or IWCV eigenmodes.  Also,  the  decay rates of
 the two least damped modes are  underpredicted by a  factor $O(10)$ in the case of the molecular viscosity parameterization,
       and  overpredicted by a factor of $O(10)$  in the case of the  eddy viscosity parameterization that produces
        the Reynolds-Tiederman turbulent mean flow. These results suggest caution 
        in interpreting the eddy viscosity required to produce the observed mean 
        flow to be an equivalent diffusion that is at the same time acting on smaller scale  structures. 
        This is consistent with the findings of Russo and Luchini  \cite{Russo-Luchini-2016}.  
        
\begin{figure*}
       \includegraphics[width = 0.75\textwidth]{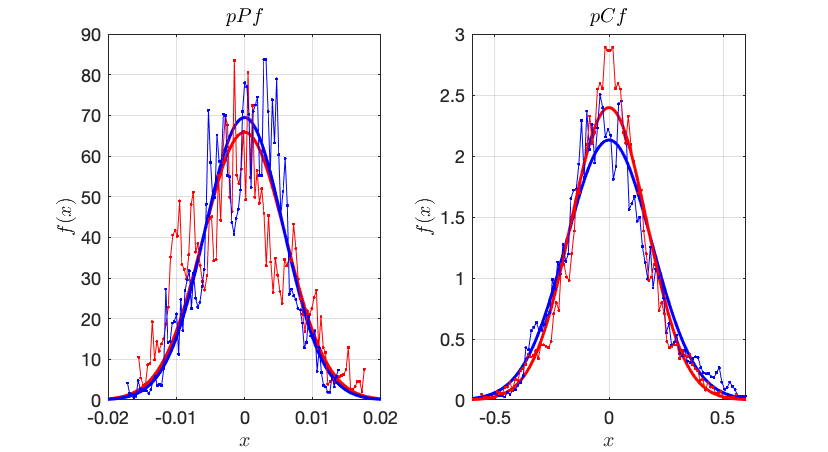}
        \caption{  Comparison between the LIM prediction  and the observed PDF of fluctuations. 
        The predicted PDF was obtained from the  $2 \times 2$ LIM. 
        Left panel: for the pPf  shown separately are the probability of projection on the symmetric POD (red) and  the antisymmetric POD (blue). The symmetric mode  has 
        $\omega=-0.05$ and $D=1.8e-06$ and the antisymmetric has $\omega=-0.03$ and $D=0.85e-06$. 
         Right panel: for the  pCf, the symmetric mode has $\omega=-0.02$ and   $D=0.6e-03$ and the antisymmetric mode has $\omega=-0.06$ and $D=1.85e-03$. }\label{fig:fokker}
 \end{figure*}

        LIM  not only predicts the decay rate of perturbations to the time-mean flow but also predicts the variance of the perturbations 
         around the decaying trajectory.  The Langevin dynamics of the fluctuations obtained using  LIM
        predicts a slight non-normal modification of the exponential decay of the modes  to the time-mean  opposed 
        by diffusive spreading of the modes away from the time-mean
        produced by the stochastic forcing.
        The resulting PDF  for the case of the $2 \times 2$ dynamics obtained by LIM, which is normal implying no mode coupling,  can be obtained from the companion Fokker-Planck equation:
%
\begin{equation}
        \frac{\partial f}{\partial t} =-\omega \frac{\partial (x f)}{\partial x} +D \frac{\partial^2 f}{\partial x^2}  ~,
\end{equation}
in which $f$ is the  perturbed mode probability distribution,  $x$ is the value of the coefficient of the projection on the associated POD mode,
and $D$ is the diffusion coefficient equal to half   
the corresponding diagonal element  of the  covariance, $\Xi$, that excites the associated eigenmode of $A$ with stable real eigenvalue, $\omega$ \footnote{In pPf: $D=1.8e-06$ for the symmetric mode with $\omega=-0.05$
and $D=0.85e-06$ for the antisymmetric mode with $\omega=-0.03$. In pCf  $D=0.6e-03$ for the symmetric mode with $\omega=-0.02$ and $D=1.85e-03$ for the antisymmetric mode with $\omega=-0.06$.}.
The perturbed mode distribution drifts back to the equilibrium under the influence of $\omega$
around which it maintains the predicted  asymptotic distribution:
\begin{equation}
f(x)=\frac{\sqrt{-\omega}}{\sqrt{2 \pi D}} \exp \left ( \frac{\omega}{2 D} x^2 \right ) ~,~~(\omega<0).
\end{equation}       
 The probability distributions for these cases is plotted in Fig. \ref{fig:fokker}. Also shown is the PDF obtained 
 by binning  the simulation data. Clearly, even this $2\times2$  LIM dynamics obtained by projection  on the first 2 POD modes suffices  to predict  the observed distribution of the fluctuations.

%
 
 \label{sec:framework}

%


\section{Conclusions}

It is accepted that the ensemble mean state is a fixed point of the ensemble dynamics to which almost all  perturbations relax. It is also believed that
sufficiently small perturbations  relax following a stable linear dynamics. Recently IWCV obtained  from a perturbed DNS
the dominant modes underlying this stable linear dynamics using an  ensemble method.  However, the interpretation of 
ensemble dynamics is not straightforward and in this case the ensemble converges  to an unstable repeller 
of the realization dynamics  rather than to  a stable fixed point.  This is common in ensemble dynamics, although 
apparently paradoxical,  as the example of the  mass-spring with random restoring force  governed by the Mathieu equation  
discussed in the Formulation section  demonstrates.  

It is generally accepted that the ensemble mean state is stable in the sense of
linear hydrodynamic stability. However, 
when account  is taken of a set of  nonlinear instabilities identified by S3T-SSD, it is a repeller in  pPf and pCf turbulence.
In this paper we have examined the implications of  this ensemble mean flow instability for the dynamics of relaxation
to the ensemble mean state.
In the S3T SSD the mean state  is the
streamwise average,  the  fluctuations are the streamwise varying components, and the  fluctuation-fluctuation 
interactions are neglected.  A close approximation to the mean flow and the fluctuation covariance in the S3T SSD is 
obtained from a quasi-linear integration of pCf dynamics from which it is
verified that the time-mean turbulent state  is not a fixed point of the S3T dynamics. 
Moreover, even if the mean turbulent state 
is enforced to be an equilibrium, this equilibrium is S3T unstable.  

It should be noted that this result applies to the 
case of post-transitional wall-bounded turbulence and
should be contrasted to the case for pre-transitional pCf and to the  turbulence that occurs in  barotropic  or baroclinic quasi-geostrophic 
models of planetary turbulence,  in which often the  time-mean turbulent state is a 
stable fixed point of both the ensemble and the S3T SSD dynamics.
The SSD fixed point in these cases corresponds, in both the ensemble SSD and the S3T SSD,  to the 
statistical mean turbulent state, both the ensemble mean and  perturbations (up to second order Gaussian approximation in the case of the S3T SSD). 
This fixed point solution in the case of the planetary turbulence exhibits the characteristic structure
 of alternating zonal jets \cite{ Farrell-Ioannou-2008-baroclinic,Farrell-Ioannou-2009-equatorial,Farrell-Ioannou-2017-Saturn}
 together with the associated covariances which support the jets by upgradient Reynolds stresses \cite{Salyk-etal-2006}. 
 A familiar physical example of this SSD equilibrium is the zonal jets 
 of Jupiter that have been demonstrated to have a fixed structure over at least decadal periods
\cite{Ingersoll-etal-2004}.

  The difference between wall-bounded post-transitional turbulence  and its pre-transitional and planetary counterparts is that in the latter cases the
 S3T  SSD
 has an attracting fixed point which coincides
 with the ensemble/time-mean state of the turbulence and standard methods of eigenanalysis
can be used to obtain the dynamics of  perturbations back to the ensemble/time-mean state. 
In wall-turbulence the S3T trajectory lies on
a transient chaotic attractor and eigenanalysis of the S3T state can not be performed  to obtain the
perturbation dynamics.
However, nothing essential is lost as the perturbation dynamics  can be
obtained by averaging over the transient S3T attractor,  which was accomplished
using LIM analysis.
Our results lend  credence to the concept  that the fundamental attractor of the 
turbulent state is identified by  the S3T attractor in cumulant variables. When this  attractor is visualized  in velocity variables   it 
corresponds to a turbulent state with the mean state 
containing  the majority  of the energy accompanied  by  a multivariate elliptical PDF of fluctuations in its immediate vicinity accounting for the remaining energy. 
Similarly, if we visualize the DNS attractor in velocity variables 
we find the majority of the energy to be contained in the mean flow with fluctuations in the vicinity accounting for the remaining energy.
The S3T attractor identifies the structure and dynamics of this fundamental entity of the turbulence as comprising
the first and second cumulants as the fundamental variables, visualized as the mean flow and the surrounding elliptical perturbation  distribution,
and their interaction as
controlling the evolution of the turbulent state in the fundamental SSD variables or equivalently as visualized in velocity variables. 
Therefore the S3T SSD predicts both the underlying structure and  evolution of the turbulent state in phase space.

%
%
%
%


Similarity 
between the dynamics of perturbations to the ensemble mean state in  S3T SSD simulations,  in which  
the structure and dynamics on the SSD attractor have been explicitly identified,  and
 DNS suggests that the existence of the SSD transient attractor and the concept of  averaging over 
 this  attractor in order to identify the dynamics 
 of  perturbations to the statistical time-mean state extends to DNS. 
 These results suggest that the essentially complete characterization of turbulence in the S3T SSD, 
 and especially the identification of a chaotic attractor in the phase space of the statistical state variables,
 can be profitably exploited to gain further insight into the  dynamics of NS turbulence.



%
 
 \section{Appendix: The structure of the  relaxation dynamics produced by  LIM  on 6 POD modes }
When the mean-flow velocity fluctuations are projected on the top 6 POD modes ordered according to their contribution 
to the mean fluctuation energy, LIM identifies that the fluctuations are governed by operator:
 \begin{eqnarray*}
A_6=  \left(
\begin{array}{cccccc}
   -0.021  & 0  & 0  & 0.048&0  & -0.117\\
  0 & -0.023  & -0.040  &0 &-0.004  & 0\\  
  0 & 0.024  & -0.050  &0 &0.044  & 0\\  
   -0.035  & 0  & 0  &-0.076 &0  & 0.200\\
    0 & -0.043  & -0.038  &0 &-0.091  & 0\\ 
  0.024  & 0  & 0  &- 0.046 &0  & -0.239  
\end{array}
\right)~,
\end{eqnarray*}
and are excited by a temporally delta-correlated white noise forcing
with spatial covariance:
\begin{eqnarray*}
\Xi_6 = 10^{-5} \left(
\begin{array}{cccccc}
   0.177  & 0  & 0  & 0.036 &0  & 0.013\\
  0 & 0.15  & 0.04  &0 &0.15  & 0\\  
  0 & 0.04  & 0.302  &0 &0.05  & 0\\  
   0.036 & 0  & 0  &0.34 &0  & -0.094\\
    0 & 0.15  & 0.05  &0 &0.26  & 0\\ 
  0.013  & 0  & 0  &- 0.094 &0  & 0.475  
\end{array}
\right)~.
\end{eqnarray*}
Note that   the POD modes, when ordered in variance, have the  following symmetry: s,a,a,s,a,s  respectively (``s'' denotes symmetry about the $x-z$ center plane at $y/h=1$ and
``a'' antisymmetry). Because of the statistical symmetry of the fluctuations, interactions among them are restricted 
to the respective symmetric and antisymmetric subspaces, and consistently the only non-zero entries in the $A_6$ and $\Xi_6$ matrices
are either in the coordinates of the symmetric modes corresponding to the POD modes  1,4,6  or those of the antisymmetric  
corresponding  to the POD modes 2,3,5 .

That the  dynamics is non-normal in the energy inner product can be ascertained by calculating the projections among the eigenmodes
indicated by the matrix of the inner product, $u^\dagger u$: 
\begin{eqnarray*}
u^\dagger u = \left(
\begin{array}{cccccc}
   1  & 0   & 0.3+0.3i &0.3-0.3i  &0& 0\\
  0 & 1  & 0  &0 &0.7+0.05 i  & 0.7-0.05i\\  
  0.3-0.3i & 0  & 1  &0.3-0.3i &0  & 0\\  
    0.3+0.3i   & 0  & 0.3+0.3i  &1 &0  &0  \\
    0 & 0.7-0.05i  & 0  &0 & 1  &0.9+0.05i\\ 
  0  & 0.7+0.05i  & 0 &0 & 0.9-0.05i  & 1 
\end{array}
\right)~.
\end{eqnarray*}
Here the eigenmodes are ordered  increasing in decay rate.  The symmetric eigenmodes  are  2,5,6 and the antisymmetric are 1,3,4 and 
it can be seen,  as  expected by the statistical symmetry of the dynamics, that non-normal interaction is confined to the mutually orthogonal symmetric and antisymmetric subspaces.

%

Non-normality of $A_6$  and non-commutation between $A_6$ and $\Xi_6$ results in  the POD modes of the Langevin dynamics differing from the  eigenmodes of the operator of the dynamics,
$A_6$,   as discussed by  \citet{North-84,North-etal-2009}.
However, the statistical symmetry of the dynamics still constrain the POD modes to lie in the mutually orthogonal  symmetric and the antisymmetric subspaces of the eigenmodes.
The commutator between $A_6$ and $\Xi_6$ is:
\begin{eqnarray*}
[ A_6, \Xi_6] = \left(
\begin{array}{cccccc}
   0.001  & 0   & 0 &0.021  &0& -0.044\\
  0 & 0.003  & 0.005  &0 &0.006  & 0\\  
  0 & 0.004  & 0.007  &0 &0.004  & 0\\  
    0.009   & 0  & 0  &-0.026 &0  &0.015  \\
    0 & -0.008  & 0.006  &0 & -0.01  &0\\ 
  -0.015  & 0 & 0 &0.022 & 0  & 0.025 
\end{array}
\right)~.
\end{eqnarray*}
%
 The requirement for identity of the POD modes with the modes of the operator in general Langevin dynamics can be immedialtely seen as follows: the POD modes are the 
 eigenmodes of the fluctuation covariance $C$, which is
 given in Langevin systems by $\int_0^\infty dt ~e^{At} \Xi e^{A^\dagger t}$. Therefore, the
eigenmodes of $C$ coincide with the eigenmodes of $A$ if $[A,A^{\dagger}]=0$, i.e. $A$ is normal, and $[A,\Xi]=0$, i.e. the forcing does not mix the eigenmodes of $A$. 

 \begin{acknowledgments}
 We thank Dr. Marios-Andreas Nikolaidis  for making available to us  the pPf  DNS data.
\end{acknowledgments}

%

\end{document}